\documentclass[11pt]{article}
\RequirePackage[margin=1in]{geometry}
\RequirePackage{amsthm,amsmath,amsfonts,amssymb}
\RequirePackage[authoryear]{natbib}
\setcitestyle{round}
\RequirePackage{graphicx}
\RequirePackage{xcolor}
\RequirePackage{booktabs}
\RequirePackage{multirow}
\RequirePackage{array}
\RequirePackage{mathtools,mathrsfs,bm}
\RequirePackage{dsfont}
\RequirePackage{enumitem}
\RequirePackage{subcaption}
\RequirePackage{tikz}
\RequirePackage{authblk}
\usetikzlibrary{cd,decorations.pathmorphing}
\RequirePackage[colorlinks,citecolor=blue,urlcolor=blue,linkcolor=blue]{hyperref}
\newcommand{\doi}[1]{\href{https://doi.org/#1}{https://doi.org/#1}}
\RequirePackage[capitalise,nameinlink]{cleveref}
\setlist[enumerate]{noitemsep}\setlist[description]{noitemsep}\setlist[itemize]{noitemsep}
\theoremstyle{plain}\newtheorem{theorem}{Theorem}\newtheorem{prop}{Proposition}\newtheorem{lemma}{Lemma}
\theoremstyle{definition}\newtheorem{definition}{Definition}\newtheorem{remark}{Remark}\newtheorem{example}{Example}
\DeclareMathAlphabet{\mathpzc}{OT1}{pzc}{m}{it}
\let\E\relax\DeclareMathOperator\E{\mathsf{E}}\let\P\relax\DeclareMathOperator{\P}{\mathsf{P}}
\DeclareMathOperator{\Var}{\mathsf{Var}}
\newcommand{\R}{\mathbb{R}}\newcommand\independent{\protect\mathpalette{\protect\independenT}{\perp}}\def\independenT#1#2{\mathrel{\rlap{$#1#2$}\mkern2mu{#1#2}}}
\makeatletter\providecommand{\leftsquigarrow}{\mathrel{\mathpalette\reflect@squig\relax}}\newcommand{\reflect@squig}[2]{\reflectbox{$\m@th#1\rightsquigarrow$}}\makeatother
\newcommand\PH{\mathrm{PH}}\newcommand\OR{\mathrm{OR}}\newcommand{\EH}{\mathrm{EH}}\newcommand{\EB}{\mathrm{EB}}\newcommand{\yt}{\tilde{Y}}\newcommand{\ytl}{\tilde{y}}

\begin{document}
\title{Counterfactual Optimization of Policy Interventions: Lexical Ordering and Leapfrogging}
\author[1]{Martina Scauda}
\author[2]{Tobias Freidling}
\author[1]{Qingyuan Zhao}
\affil[1]{Statistical Laboratory, University of Cambridge, \texttt{ms2985@cam.ac.uk}, \texttt{qyzhao@statslab.cam.ac.uk}}
\affil[2]{Institute of Mathematics, École Polytechnique Fédérale de Lausanne, \texttt{tobias.freidling@epfl.ch}}
\date{}
\maketitle
\begin{abstract}
Most data-driven policy learning methods maximize average
outcomes, overlooking the possibility that a policy beneficial on average may
still harm a substantial fraction of individuals. Motivated by the
ethical principle of ``first do no harm'',  we study how to design a
change from a baseline policy that improves overall welfare while
keeping the worst-case probability or expectation of individual harm
below a specified limit. We establish sufficient conditions under
which an optimal policy transition has a \emph{lexical leapfrogging}
structure: groups defined by covariates and current treatment are
ranked by a priority score, and any treatment change moves them
directly to the conditionally optimal treatment. We derive this score
under several models for the dependence among potential outcomes.
We demonstrate this harm-aware policy
optimization approach in a reanalysis of the I-SPY2 breast cancer
platform trial and show how the consideration of counterfactual harm
may lead to different conclusions about which treatment-subgroup pairs
may warrant deprioritization in further clinical evaluation.
\end{abstract}
\section{Introduction}

The Hippocratic principle of \textit{Primum non nocere} (``first do
no harm'') establishes a moral imperative in clinical practice to weigh
the duty of non-maleficence against the goal of beneficence
\citep{jonsen1978no}. However, contemporary data-driven algorithms for
individualized treatment rules or policies are typically designed to
optimize population benefits and ignore the ``no harm''
consideration. In the case of a one-off binary intervention, this
leads to the optimal policy that simply assigns treatment based on the sign of
the conditional average treatment effect (CATE)
\citep{murphy_optimal_2003}.
With the rise of black-box decision systems, a growing debate in
public health and policy research concerns how considerations of harm
can be incorporated into policy learning beyond the CATE
\citep{li19_unit_selec_based_count_logic,mueller_personalized_2023,dawid23_person_decis_makin_without_count,sarvet23_persp_harm_person_medic,benmichaelpolicy2024,sawant26_count_harm}.

One of the fundamental difficulties in applying the ``no harm'' principle
to policy learning is in understanding the trade-off between overall
population benefits and potential individual harm. A numerical example in
\citet{mueller_personalized_2023} has been intensively debated in the literature: a
treatment diminishes the death rate by the same 28\% in both women and
men, yet the reduction in women's death is composed entirely of women who were
cured by the treatment and would have died if left untreated, whereas
the reduction in men's death is split between 49\% who are saved and
21\% who are harmed (would have survived if untreated). While it is
uncontroversial that women should always receive
treatment, this example also raises the question of whether there is any
ethical ground, given evidence of average benefit but potential
individual harm, to not give the treatment to men.\footnote{
A peculiarity of the
  numerical example in \citet{mueller_personalized_2023} is that the unconditional
probabilities of counterfactual harm ($0\%$ for women and $21\%$ for
men) are point identified with the help of an additional
non-experimental dataset. In
reality, counterfactual harm is rarely point identified. See
\citet{sarvet25_rejoin_to_person_medic_alter_persp,dawid23_person_decis_makin_without_count}
for critiques of the numerical example and its implications.}
Without further information, for example, on the natural treatment
value \citep{sarvet25_rejoin_to_person_medic_alter_persp}, this appears
to be a very difficult ethical question to resolve at the individual
level.

This article contributes to this debate by providing a different,
population-based perspective and some concrete solutions. From a
policy maker's perspective, it is perhaps inevitable that an
intervention, with all the good intentions, inflicts some individual
harm; that is, it is often unrealistic to demand that a
population-based intervention should cause no harm at all. What the
policy maker could do, instead, is to maximize the societal benefits
while making sure that the potential harm is controlled at an
acceptable level. Thus, attention should be paid to which groups
should be \emph{prioritized} to receive the intervention; indeed, in
the above numerical example, we believe it is uncontroversial to say
that, without further information, women should receive the treatment
before men if not everyone can be treated (e.g.\ due to resource
constraints).

Our work advances the emerging literature on harm-aware policy
learning, reviewed in \cref{sec:related-work}, in several
directions. We study policy \emph{transitions} from a baseline policy
already in use, extending earlier formulations
\citep{richens2022counterfactual,kallus2022} to
stochastic baselines, multi-valued treatments, and optimal
transition kernels. This framework is natural in clinical and social
policy settings, where there is often a status quo to modify. More
fundamentally, specifying a reference policy is necessary because harm
is inherently relative. Our key finding, which emerges when moving
from binary to multiple treatment arms, is that an optimal policy
transition often has a \emph{lexical leapfrogging} structure:
subgroups, defined by their covariates and current treatment, are
ranked by a priority score and moved directly — or allowed to
``leapfrog'' — to their conditionally optimal treatment. The priority
score is the CATE gain from this move divided by its conditional
worst-case expected harm, and we obtain closed-form expressions under
several models for dependent potential outcomes. This
lexical structure has a formal affinity with Rawls's lexical
difference principle and with leximin social orderings in
social-choice theory
\citep{sen2017collective,rawls1971theory,Fleurbaey_Maniquet_2011}; we
discuss the scope and limitations of this analogy in
\cref{sec:discussion}.

\subsection{Preview of the mathematical results}
\label{sec:overview-main-result}

To preview our main results, we briefly introduce some notation for
the harm-aware policy optimization problem considered here. We use a
random variable $X \in \mathcal{X}$ to denote the
background information or covariates of an individual, a real-valued
variable $Y \in \mathbb{R}$ to denote the outcome or utility of the
same individual, and a categorical $Z \in \mathcal{Z}$ ($|\mathcal{Z}|
< \infty$)
to denote the treatment variable set by the policy maker to that
individual. We adopt the Neyman-Rubin potential outcomes framework
\citep{imbens2015causal} and use $Y(z)$ to denote the potential
outcome of $Y$ when $Z$ is set to $z$ and $Y(\cdot) = (Y(z): z \in
\mathcal{Z})$ to denote the potential outcomes schedule.
By \emph{policy} we mean a conditional distribution $d(z
\mid x)$ that describes the probability of assigning $Z$ to $z$ given
$X = x$; policies are \emph{deterministic} if for all $x$, there
exists a treatment $z(x)$ such that $d(z(x) \mid
x) = 1$. We will maintain the following assumptions:
\begin{description}
\item[A1] $Z \independent Y(\cdot) \mid X$, so the policy
  assigns treatment at random given $X$ (this is always true if the
  policy is deterministic);
\item[A2] The conditional distribution of $Y(z)$ given $X = x$, denoted
  by $\gamma_z(x)$, is given (e.g., estimated from randomized
  experiments), and $\E\{\lvert Y(z)\rvert\}<\infty$.
\end{description}
Let $\gamma(x)$ denote the conditional distribution of $Y(\cdot)$ given
$X = x$. Let $\Gamma_{\mathrm{F}}(x)$ denote the Fr\'{e}chet class
\citep[after][]{frechet1951tableaux} that
contains all joint distributions $\gamma(x)$ that have the given marginals
$\gamma_z(x)$ for all $z \in \mathcal{Z}$ given $X=x$ and $\Gamma_{\mathrm{F}} =
\prod_{x \in \mathcal{X}} \Gamma_{\mathrm{F}}(x)$.

Without any further constraint, the optimal policy that maximizes the
expected value of $Y(Z)$ simply assigns $Z$ to the treatment value
that maximizes the conditional potential outcome expectation; that is,
assuming no ties, the optimal policy is given by
\[
  d(z^{*}(x) \mid x) = 1,~\text{where}~z^{*}(x) = \arg\max_z
  \E\{Y(z) \mid X = x\}.
\]
There exists a large literature in statistics on estimating this optimal
policy \citep[see
e.g.][]{qian2011performance,kosorok2019precision}. Here we
consider a more general perspective centered on policy transitions.

\begin{definition}
  A \emph{policy transition} $\kappa$ is a Markov kernel from the
  current treatment $Z$ to a new treatment $Z'$ conditional on $X$,
  represented by the
  transition probabilities $\kappa(z' \mid z, x) \geq 0$ such that
  $\sum_{z' \in \mathcal{Z}}
  \kappa(z' \mid z, x) = 1$ for all $z \in \mathcal{Z}$ and $x \in
  \mathcal{X}$.
\end{definition}

In this paper, we focus on the following harm-constrained policy
optimization problem:
\begin{equation}
  \label{eq:opt-eh}
  \underset{\kappa}{\text{maximize}} \quad \E_{\kappa} \{ Y(Z') \} \quad
  \text{subject to} \quad \sup_{\gamma \in \Gamma} \EH_{\kappa,\gamma} \leq B,
\end{equation}
where the following assumptions will be maintained throughout the paper:
\begin{description}
\item[A3] $Z'$ is the new treatment generated by $Z' \mid Z, X \sim
  \kappa$ at random, meaning $(Z,Z') \independent Y(\cdot) \mid X$;
\item[A4] $\Gamma \subseteq \Gamma_{\mathrm{F}}$ is a given product subclass of the
  Fr\'{e}chet class, meaning $\Gamma =
  \prod_{x \in \mathcal{X}} \Gamma(x)$ for some $\Gamma(x) \subseteq \Gamma_{\mathrm{F}}(x)$;
\item[A5] $\EH_{\kappa,\gamma} = \E_{\kappa,\gamma}\{h(Y(Z') - Y(Z))\}$ is the
  \emph{expected harm} of the transition $\kappa$, where
  $h: \R \to \R_{\geq 0}$ is a measure of individual harm that
  satisfies $h(y) = 0$ for all $y \geq 0$;
\item[A6] $B \geq 0$ is given.
\end{description}
Assumptions A1--A6 define an oracle decision model in which the
potential outcomes marginal laws and the dependence class $\Gamma$ are taken as
given. Identification of the marginal laws from observed data
additionally requires consistency and no interference, conditional
exchangeability given sufficient covariates and design history, and
positivity among eligible arms, followed where necessary by
standardization to the target distribution of $X$ \citep{imbens2015causal,hernan_robins_2020}.

In words, \eqref{eq:opt-eh} tries to find the optimal policy transition
$\kappa$ that maximizes the average utility subject to a constraint on
the expected harm that might be inflicted in the worst case.
As a concrete example, one can take $h$ to be the negative part
function, $h(y) = \max\{-y, 0\}$; when $Y$ is binary, the
expected harm then reduces to the \emph{probability of harm}
\citep{mueller_personalized_2023}: $\EH_{\kappa,\gamma} =
\P_{\kappa,\gamma}(Y(Z') < Y(Z))$.

Let $\kappa^{*}$ denote the solution to \eqref{eq:opt-eh}; here
we assume no ties and omit the dependence of $\kappa^{*}$ on $\Gamma$
and the distribution of $(X,Z)$. A key finding in this paper is:
\begin{center}
\textbf{The optimal policy transition $\kappa^{*}$
  often has a ``lexical
  leapfrogging'' structure.}
\end{center}
By \emph{lexical}, we mean there exists a
priority score $\Lambda: \mathcal{Z} \times \mathcal{X} \to [0, \infty]$ such
that the following proposition is true:
  \begin{align}
    % \label{eq:lexical-1}
    % \kappa^{*}_B(z \mid z, x) = 1 \implies \kappa^{*}_{B'}(z \mid z,
    % x) = 1 \quad \text{for all}~z, x, B' < B.
    % \\
    \label{eq:lexical-2}
    \kappa^{*}(z \mid z, x) = 1,~\Lambda(z',x') < \Lambda(z, x)
  \implies \kappa^{*}(z' \mid z',
                                    x') = 1 \quad \text{for
    all}~z,x,z',x'.
  \end{align}
% Equation \eqref{eq:lexical-1} says no transition out of $(z,x)$
% implies no such transition under a stricter harm constraint.
This equation means no transition out of $(z,x)$ implies no
transition out of any pair $(z',x')$ of lower priority. By
\emph{leapfrogging}, we mean
\begin{equation}
  \label{eq:leapfrogging}
  \kappa^{*}(z' \mid z,x) = 0\quad\text{for all}~z',z,x~\text{such
    that}~z' \not \in \{z, z^{*}(x)\}.
\end{equation}
That is, transitions only occur to the optimal treatment at the
subgroup level determined by $X$. % Moreover, the optimal policy transition $\kappa^{*}_B$ is
% independent of the initial policy.

The next result is a corollary of \Cref{thm:lexical-leapfrogging}
below and exemplifies this phenomenon.

\begin{theorem} \label{thm:leapfrogging-binary}
Suppose the outcome variable $Y$ is binary and
$\Gamma=\Gamma_{\mathrm F}$. Then there exists an optimal lexical
leapfrogging policy transition whose priority score is independent of
the initial policy $d$.
\end{theorem}

% \subsection{A numerical demonstration of lexical leapfrogging}
% \label{sec:demontr-lexic-leapfr}

\begin{figure}[tbp]
  \centering

  %------------------------------------------------------------
  % Panel (a): DGP, scores, and rankings
  %------------------------------------------------------------
  \begin{subfigure}[t]{0.8\linewidth}
    \centering
    \small
    \setlength{\tabcolsep}{5pt}

    \begin{tabular}{cc|c|rr|rr|rr}
      \toprule
      & &
      \multicolumn{1}{c|}{} &
      \multicolumn{2}{c|}{CATE} &
      \multicolumn{2}{c|}{$\Lambda_1$ (Non-neg.\ dep.)} &
      \multicolumn{2}{c}{$\Lambda_0$ (Fr\'{e}chet)} \\
      $x$ & $z$ & $\gamma_z(x)$
          & $\tau_{z^*,z}$ & Rank
          & Score & Rank
          & Score & Rank \\
      \midrule

      % \multirow{3}{*}{$G_1$}
      $G_1$
        & $0$   & $0.15$ & $0.50$ & $1$
        & $9.524$ & $\mathbf{2}$
        & $3.333$ & $\mathbf{3}$ \\
        & $1^*$ & $0.65$ & \textemdash & \textemdash
        & \textemdash & \textemdash
        & \textemdash & \textemdash \\
        & $2$   & $0.20$ & $0.45$ & $2$
        & $6.429$ & $\mathbf{4}$
        & $2.250$ & $\mathbf{4}$ \\
      % \midrule
      % \multirow{3}{*}{$G_2$}
      $G_2$
        & $0$   & $0.55$ & $0.40$ & $3$
        & $14.545$ & $\mathbf{1}$
        & $8.000$ & $\mathbf{1}$ \\
        & $1$   & $0.65$ & $0.30$ & $4$
        & $9.231$ & $\mathbf{3}$
        & $6.000$ & $\mathbf{2}$ \\
        & $2^*$ & $0.95$ & \textemdash & \textemdash
        & \textemdash & \textemdash
        & \textemdash & \textemdash \\
      \bottomrule
    \end{tabular}

    \caption{Data-generating process ($\gamma_z(x) = \P(Y=1
      \mid Z=z, X=x)$) and rankings under CATE and the
      harm-aware scores. Bold ranks differ from the CATE ranking;
      $z^*$ denotes the conditionally optimal arm.}
    \label{fig:leapfrog-rankings}
  \end{subfigure}

  \vspace{0.8em}

  %------------------------------------------------------------
  % Panel (b): Value--harm trajectories
  %------------------------------------------------------------
  \begin{subfigure}[t]{0.8\linewidth}
    \centering
    \includegraphics[width=0.75\linewidth]{
      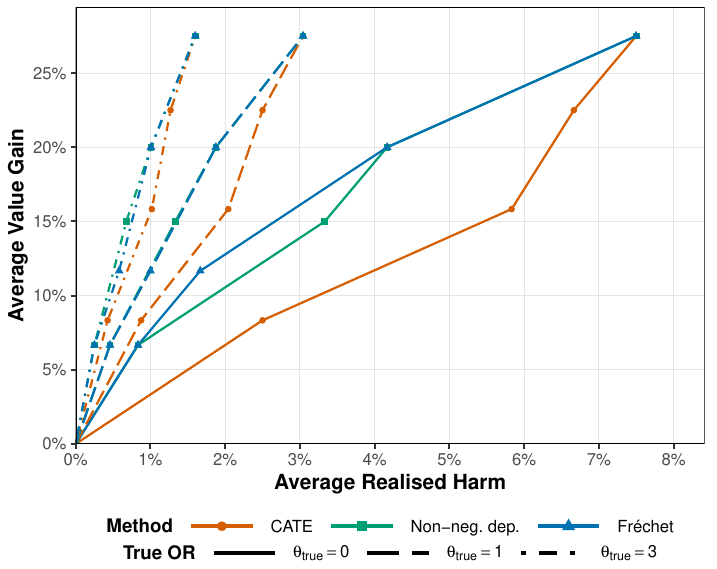
    }
    \caption{Cumulative value gain against realized harm when
      group--arm pairs are switched in the order induced by each
      score, for different values of the true odds ratio
      $\theta_{\mathrm{true}}$.}
    \label{fig:leapfrog-pareto}
  \end{subfigure}

  \caption{A demonstration of harm-aware policy learning and the
    lexical leapfrogging structure in a numerical example with two
    covariate groups and three treatment arms.}

  \label{fig:leapfrog-combined}
\end{figure}

We demonstrate the lexical leapfrogging mechanism in a numerical
example with three treatment arms $Z\in\{0,1,2\}$, a binary outcome $Y
\in \{0,1\}$, and two covariate groups. For both groups, arm $0$ is
the worst, but their optimal treatment arms differ. We consider three
ways of ranking the $6$ treatment-covariate subgroups $(z,x)$ that are
assumed to have equal probabilities:
\begin{enumerate}
\item using the CATE of switching from arm $z$ to
the optimal arm, i.e.\ $\tau_{z^*,z}=\E[Y(z^*)\mid X] -
\E[Y(z)\mid X]$;
\item using the harm-aware priority score $\Lambda_1$ calculated under
  the assumption of non-negatively correlated potential outcomes;
\item using the harm-aware priority score $\Lambda_0$ calculated under
  the Fr\'{e}chet class $\Gamma_{\mathrm F}$.
\end{enumerate}
Closed-form formulas for $\Lambda_0$ and $\Lambda_{1}$ can be found in
\Cref{sec:priority-binary}. \Cref{fig:leapfrog-rankings} shows a conditional distribution of $Y$
given $Z$ and $X$ and the rankings obtained by these three approaches,
which are all different. \Cref{fig:leapfrog-pareto} plots the expected
value increase of the policy changes based on these different orders
against the realized harm under different values of the odds ratio between the
potential outcomes, showing CATE-based ranking can be far from the
Pareto frontier.

\subsection{Related work and our contributions}\label{sec:related-work}

A broad and increasingly interconnected literature studies how to learn
treatment-assignment rules from experimental or observational data. This
problem appears under several names, including statistical treatment choice
and empirical welfare maximization in econometrics, and individualized
treatment rules, optimal treatment regimes, and policy learning in statistics,
biostatistics, and machine learning
\citep{manski_statistical_2004,murphy_optimal_2003,
qian2011performance,kitagawa_who_2018,athey_policy_2021,
kosorok2019precision}. These formulations generally seek a policy with high
population value, and their statistical performance is commonly
evaluated through value or welfare regret.

Constrained policy optimization arises naturally under scarce resources
or heterogeneous treatment costs. With binary treatments, linear
resource constraints often yield
rank-and-treat rules based on a benefit-to-cost score
\citep{bhattacharya_inferring_2012,luedtke_optimal_2016,
sun_treatment_2021,qiu_individualized_2022}. Closely related,
\citet{levis_intervention_2024} prioritize whom to contact, leaving
uncontacted individuals at their natural treatment and assigning contacted
individuals their conditionally optimal treatment. With multiple costly arms,
\citet{sverdrup_qini_2025} characterize allocation using a cost--gain convex
hull, allowing arms below the hull to be skipped. These methods
produce allocation rules resembling ours, but impose different
constraints: they limit resource use or contact from a utilitarian
perspective, whereas ours limits the counterfactual harm motivated by
ethical considerations.

Several recent papers incorporate such counterfactual comparisons directly
into decision making. \citet{richens2022counterfactual} define harm relative
to a default action or policy and construct harm-averse objectives.
\citet{kallus2022} derives sharp bounds for the probability of
harm. \citet{li_trustworthy_2023} learn binary-treatment policies that
maximize reward subject to an upper bound on the fraction counterfactually
harmed. \citet{benmichaelpolicy2024} study utilities that value benefit and
harm asymmetrically, derive minimax policies under partial identification,
and also consider a constrained harm-and-resource
formulation. In related work, \citet{koch2025counterfactual} show
that counterfactual-risk differences are point identified precisely when the
loss is additive across potential outcomes.
In the work most closely related to ours, \citet{wu_safe_2025} study similar
harm-constrained optimization problems with a binary treatment, a
binary outcome, and a control-only baseline policy.

The counterfactual harm constraint makes the optimization problem
\eqref{eq:opt-eh} a decision problem under
partial identification \citep{manski_identification_2007}. This is
because randomization or conditional ignorability identifies the
conditional marginal distributions of the potential outcomes, but not
their within-individual coupling. Sharp
fixed-marginal bounds draw on classical Fr\'{e}chet and extremal-coupling theory
\citep{frechet1951tableaux,cambanis_inequalities_1976,
tchen_inequalities_1980,makarov_estimates_1982} and have been applied to
treatment-effect distributions and the probability of benefit or harm
\citep{heckman_making_1997,fan_park_2010,
tian_probabilities_2000,kallus2022}. Sensitivity analyses instead restrict the
unidentified dependence between potential outcomes using odds ratios, departures from conditional
independence, or potential-outcome correlations
\citep{zhang_heterogeneity_2013,ciocanea_teodorescu_2025,
wu2026individualrisk}.

Relative to the existing literature, our conceptual contribution is a
substantially more general framework for harm-aware policy learning: it
accommodates transitions from a possibly stochastic incumbent policy,
multi-valued treatments, transition decisions depending on
both covariates and the current assignment, general harm measures, and
alternative restrictions on the unidentified dependence between
potential outcomes. Our main technical contribution is the
leapfrogging lemma in \Cref{sec:multi-valu-treatm}: under
subadditivity of harm and appropriate
stochastic-dominance and recoupling conditions, a policy transition
through intermediate treatments can be replaced without reducing
average outcome or increasing worst-case harm by a leapfrogging
change. This nontrivial lemma then allows us to reduce the
multi-valued treatment problem to a continuous knapsack problem and
obtain a lexical solution.

\subsection{Paper Outline}
\label{sec:structure}
The rest of the paper is organized as follows. In
\Cref{sec:optimization-bin}, we consider the counterfactual policy
optimization problem \eqref{eq:opt-eh} with a binary treatment and
discuss how lexical orderings arise from constrained optimization
problems. In \Cref{sec:multi-valu-treatm}, we consider the
multi-valued treatment problem and discuss why the solution often has
a leapfrogging property by the classical Neyman-Pearson lemma. In
\Cref{sec:leapfr-prior}, we combine these
insights and prove a generalization of
\Cref{thm:leapfrogging-binary}. In \cref{sec:priority}, we extend the
previous result to general outcomes and provide a characterization of
the priority score under additional assumptions on the dependence
between potential outcomes. In \cref{sec:numerical}, we illustrate the
validity of our method using simulations calibrated to the data
adopted for the subsequent real-world analysis. We conclude with a
discussion on the connection to distributive justice and future
research questions in \cref{sec:discussion}. Most technical proofs are
deferred to the Appendix.

\section{Binary treatment and
  lexical ordering}\label{sec:optimization-bin}

\subsection{A review of policy optimization with binary treatments}
\label{sec:revi-policy-optim}

In this section, we focus on the policy learning problem with a binary
treatment $Z \in \mathcal{Z} = \{0,1\}$ that is intensively studied in
statistics and econometrics
\citep{kosorok2019precision,athey_policy_2021}. We mainly use this
setting to demonstrate how lexical ordering naturally arises from
solving continuous knapsack problems
\citep{NeymanPearsonLemma,dantzig1957}.

Let us first briefly review the policy learning literature in this
case that focuses on finding optimal policies (instead of policy
transitions). With a binary treatment, it is often simpler
to represent the policy by
\[
  d(x) = \P(Z = 1 \mid X = x)
\]
and consider
the conditional average treatment effect (CATE)
\[\tau(x) =
\E\{Y(1) - Y(0) \mid X = x\}\] instead of the individual potential
outcome means. Without any other constraints, it is easy to see that
\[
  z^{*}(x) = \arg\max_z \E\{Y(z) \mid X = x\} = \mathds{1}_{\{\tau(x) > 0\}},
\]
assuming $\tau(x) \neq 0$ so that the solution is unique. In other words, the
optimal unconstrained policy simply uses the sign of the CATE
\citep{murphy_optimal_2003}.

In practice, the feasible policy class may be limited by a resource
constraint as in the optimization problem below:
\begin{align} \label{eq:resource-constraint}
  \max \E_d\{Y(Z)\} \quad \text{subject to}~\E\{d(X) w(X)\} \leq B,
\end{align}
where $w(X)$ is the cost of switching a unit with covariate $X$ from
$Z = 0$ to $Z = 1$, and $B$ is a given budget. (This
common setup assumes the baseline policy to be $Z = 0$ with probability~$1$.) By expanding the objective function as $\E_d\{Y(Z)\} =
\E\{Y(0)\} + \E\{d(X) \tau(X)\}$, it is easy to see that
\eqref{eq:resource-constraint} is an instance of the continuous
knapsack problem and admits a greedy/lexical solution
\citep{NeymanPearsonLemma,dantzig1957}. The following result is a
restatement of the celebrated Neyman-Pearson lemma and a proof can be
found in \citet[p.~66]{LehmannRomano}.

\begin{lemma}\label{lem:np}
    Let $X$ be a given random variable and let 
    $a,b: \mathcal{X} \to \R_{\geq 0}$ satisfy
    $\E\{a(X)\}<\infty$ and $\E\{b(X)\}<\infty$. For $B\geq0$,
    consider the optimization problem
    \begin{equation}
    \underset{c: \mathcal{X} \to [0,1]}{\max} \E \{a(X)
    c(X)\}, \quad
    \text{subject to } \E\{b(X) c(X)\} \leq B.
    \label{eq:np-knapsack}
  \end{equation}
  A solution is given by
   \begin{equation}
   c^{*}(x) = t(x;\alpha, \lambda) := \begin{cases}
  1, & \text{if}~ \Lambda(x) > \lambda ; \\
  \alpha, & \text{if}~ \Lambda(x) =\lambda; \\
  0, & \text{if}~ \Lambda(x) < \lambda,
\end{cases}
       \label{eq:sol-lemma}
     \end{equation}
     where $\Lambda(x)=a(x)/b(x)$ when $b(x)>0$,
     $\Lambda(x)=+\infty$ when $b(x)=0<a(x)$, and
     $\Lambda(x)=0$ when $a(x)=b(x)=0$.
     If $B<\E[b(X)\mathds{1}_{\{a(X)>0\}}]$, the parameters can be
     chosen so that $\E[b(X)t(X;\alpha,\lambda)]=B$. If
     $B\geq\E[b(X)\mathds{1}_{\{a(X)>0\}}]$, one may instead take
     $c^*(x)=\mathds{1}_{\{a(x)>0\}}$; the budget constraint then need
     not bind.
\end{lemma}

Applying \Cref{lem:np} to the optimization problem in
\eqref{eq:resource-constraint},
we see that its optimal solution has a lexical ordering: $d^{*}$
allocates treatment to groups with the highest reweighted CATE
\[
  \Lambda(x) = \max\{\tau(x), 0\}/w(x)
\]
until resources are exhausted
\citep{bhattacharya_inferring_2012,luedtke_optimal_2016,qiu_individualized_2022,sun_treatment_2021}. Other
models of resource constraints may lead to different priority indices,
such as the conditional potential benefit that measures maximal gain
relative to no intervention \citep{levis_intervention_2024}.

\subsection{Harm-constrained policy optimization}

To apply the same argument to solve the harm-constrained policy
optimization problem \eqref{eq:opt-eh}, we start with the observation
that
\begin{align}
  \sup_{\gamma \in \Gamma} \EH_{\kappa,\gamma}
  =& \sup_{\gamma \in
                                                    \Gamma}
     \E_{\kappa,\gamma}\Big[\E\{h(Y(Z') - Y(Z)) \mid X\}\Big]
  = \E\Big[\sup_{\gamma \in \Gamma} \EH_{\kappa,\gamma}(X)\Big], \label{eq:sup-exp-1}
  % =& \E\left[\sup_{\gamma \in \Gamma} \E\{(1 - d(X)) \kappa(1 \mid 0,
  %   X) (Y(1) - Y(0))_- + d(X) \kappa(0 \mid 1, X) (Y(0) - Y(1))_- \mid
  %   X\}\right],
\end{align}
where $\EH_{\kappa,\gamma}(X) = \E_{\kappa,\gamma}\{h(Y(Z') - Y(Z))
\mid X\}$ is
the conditional expected harm given $X$, and the second equality in
the display follows from the assumption that $\Gamma$ is a product
class. (Recall that $\gamma$ is the
joint distribution of $Y(\cdot)$ given $X$.)
Because $Z$ is binary, we can rewrite the conditional worst-case harm as
\begin{align}
  \sup_{\gamma \in \Gamma} \EH_{\kappa,\gamma}(X) =
  \sup_{\gamma \in \Gamma} &(1 - d(X)) \kappa(1 \mid 0, X)
           \E_{\gamma}\{h(Y(1) - Y(0)) \mid X\} \nonumber \\
                                     &+ d(X) \kappa(0 \mid 1, X)
                                       \E_{\gamma}\{h(Y(0) - Y(1)) \mid X\}
                                       . \label{eq:sup-E-2}
\end{align}
We can then apply \Cref{lem:np} to obtain the following result. The key observation in
the proof, contained in \cref{app:proofs-bin}, is that although there
are two functions $\kappa(1 \mid 0, X)$ and $\kappa(0 \mid 1, X)$ to
be optimized, only one of them is active at a given value of $X$.

\begin{theorem}\label{thm:PropOptimal}
  Suppose the treatment $Z$ is binary. Then there exist $\lambda^{*}
  \geq 0$  and $\alpha^{*} \in [0,1]$ that can be determined from $B$ such
  that the following policy transition $\kappa^*$ solves the optimization
  problem \eqref{eq:opt-eh}:
\begin{enumerate}
    \item For $x \in \mathcal{X}$ such that $\tau(x) > 0$, we have
      $\kappa^*(0 \mid 1, x) = 0$ and $\kappa^*(1 \mid 0, x) = t(x; \alpha^{*}, \lambda^{*})$;
    \item For $x \in \mathcal{X}$ such that $\tau(x) < 0$, we have
      $\kappa^*(1 \mid 0, x) = 0$ and $\kappa^*(0 \mid 1, x) = t(x; \alpha^{*}, \lambda^{*})$;
    \item For $x \in \mathcal{X}$ such that $\tau(x) = 0$, we have
      $\kappa^*(1 \mid 0, x) = \kappa^*(0 \mid 1, x) = 0$;
    \item The other transition probabilities can be determined
      accordingly by $\kappa^*(0 \mid 0, x) = 1 - \kappa^*(1 \mid 0,
      x)$ and $\kappa^*(1 \mid 1, x) = 1 - \kappa^*(0 \mid 1, x)$;
    \end{enumerate}
where the priority score in defining the thresholding rule $t$ is
given by
\begin{equation} \label{eq:Lambda}
    \Lambda(x) =
    \begin{dcases}
        \frac{\tau(x)}{\sup_{\gamma \in \Gamma(x)} \E_{\gamma}\{h(Y(1) -
      Y(0)) \mid X = x\}
        % \min\{\P(Y(0) = 1 \mid X), \P(Y(1) = 0 \mid X)\}
      }, & \text{if}~\tau(x) > 0,\\
      0, & \text{if}~\tau(x) = 0, \\
        \frac{-\tau(x)}{\sup_{\gamma \in \Gamma(x)} \E_{\gamma}\{h(Y(0) -
      Y(1)) \mid X = x\}
        % \min\{\P(Y(0) = 0 \mid X), \P(Y(1) = 1 \mid X)\}
        }, & \text{if}~\tau(x) < 0.
    \end{dcases}
\end{equation}
\end{theorem}
A more specialized result assuming a control-only baseline policy was
independently obtained by \citet{egami_jin_26}.

\section{Multi-valued treatment and leapfrogging}
\label{sec:multi-valu-treatm}

We next consider the more general problem with a multi-valued treatment $Z \in
\mathcal{Z} = \{0,1,\dots,M\}$ and introduce conditions under which
``leapfrogging'' to the optimal treatment is preferred over gradual
improvements.

To simplify the exposition, we will omit the dependence on the
covariates $X$. Without conditioning on $X$, a \emph{policy} $d(z)$ is a
distribution on $\mathcal{Z}$ and a \emph{policy transition} $\kappa(z'
\mid z)$ is a Markov kernel on $\mathcal{Z}$. Recall that $\gamma_z$
is the marginal distribution of $Y(z)$, $\gamma$ is the joint
distribution of $Y(\cdot)$, and $\Gamma_{\mathrm{F}}$ contains all
$\gamma$ with matching marginals (the Fr\'{e}chet class). Denote
$\mu_z = \E\{Y(z)\}$, $z = 0,\dots,M$.

We next introduce the two technical conditions that are needed in the
main result of this section.

\begin{definition}\label{def:harm-func}
  We say $h:\R \to \R_{\geq 0}$ is \emph{subadditive} if $h(y + y') \leq h(y) +
  h(y')$ for all $y, y' \in \R$.
\end{definition}

Because the harm measurement $h$ is required to satisfy $h(y) = 0$
for all $y \geq 0$, subadditivity of $h$ implies that $h$ is
nonincreasing.

\begin{definition} \label{def:monotone-recoupling}
  We say $\Gamma \subseteq \Gamma_{\text{F}}$
  is \emph{closed under monotone recoupling at a pair}
  $(z,z^{*}) \in \mathcal{Z}^2$ if for every $\gamma \in \Gamma$,
  there exists a joint distribution $\gamma' \in \Gamma$ such that
  \begin{enumerate}
  \item the distribution of $(Y(0), \dots, Y(z - 1),
    Y(z+1),\dots,Y(M))$ is the same under $\gamma$ and
    $\gamma'$;
  \item $Y(z) \leq Y(z^{*})$ almost surely under $\gamma'$.
  \end{enumerate}
  % We say $\Gamma$ is \emph{closed under monotone recoupling} if the
  % above is true for all $(z_1,z_2)$.
\end{definition}

\Cref{def:monotone-recoupling} is introduced as a property of
$\Gamma$, but note that the existence of a non-empty $\Gamma$ that
satisfies the closure property implies that $\gamma_z$ must
be stochastically dominated by $\gamma_{z^{*}}$ (quantiles of $Y(z^{*})$ are
at least as large as the corresponding quantiles of $Y(z)$).
Conversely, by Strassen's theorem, stochastic dominance guarantees
closure when $\Gamma = \Gamma_{\text{F}}$, but need not do so for a
restricted subclass of $\Gamma_{\text{F}}$, whose additional
dependence restrictions may exclude the required recoupling
\citep{strassen65_exist_probab_measur_with_given_margin}.

\begin{lemma}[Leapfrogging] \label{lem:leapfrogging}
  Consider any $z, z^{*} \in \mathcal{Z}$, $z \neq z^{*}$. Suppose
  $\Gamma \subseteq \Gamma_{\text{F}}$ is non-empty and is closed
  under monotone recoupling at $(z,z^{*})$. Further, suppose the harm
  measurement $h$ is subadditive. Then for any given policy transition
  $\kappa$, there exists a policy transition $\kappa^{*}$ such that
  all the following are true:
  \begin{enumerate}
  \item $\E_{\kappa^{*}}\{Y(Z')\} \geq \E_{\kappa}\{Y(Z')\}$;
  \item $\sup_{\gamma \in \Gamma} \EH_{\kappa^{*}, \gamma} \leq
    \sup_{\gamma \in \Gamma} \EH_{\kappa, \gamma}$;
  \item $\kappa^{*}(z \mid z') = 0$ for all $z' \neq z$.
  \end{enumerate}
\end{lemma}

In words, this key lemma says that if a treatment $z$ is dominated
by another $z^{*}$ in the sense described in
\Cref{def:monotone-recoupling}, then there exists an optimal policy
transition that ``leaps over'' $z$ entirely.

The full proof of \Cref{lem:leapfrogging} is nontrivial and can be
found in \cref{app:proofs-multi}. Here we sketch the main arguments in the
proof. Without
loss of generality, we can assume $\mu_0 \leq \mu_1 \leq \dots \leq \mu_M$
and $z < z^{*}$ (as $\gamma_z$ is stochastically dominated by
$\gamma_{z^{*}}$). The proof consists of three consecutive
transformations of $\kappa$ that remove
all transitions into $z$ without decreasing the mean potential
outcome or increasing the worst-case counterfactual harm. Similar to the proof of
\Cref{thm:PropOptimal}, the first step removes all ``downward
transitions'' from any $\overline{z} > z$ to $z$. The second step then
removes all ``upward transitions'' to $z$ or from $z$; a careful
``compensated modification'' is needed here to
ensure no increase in worst-case harm with a subadditive $h$ (see
\Cref{ex:harm-increase} below). The third step
further removes all ``upward transitions'' to $z$ if they have not
been eliminated in the second step already; the monotone recoupling
property of $\Gamma$ and monotonicity of $h$ are used in this step to
prove the worst-case harm does not increase.

\begin{example} \label{ex:harm-increase}
  Here we provide an example that shows
  removing ``upward transitions'' to a suboptimal treatment may
  increase the worst-case harm. Consider $\mathcal{Z} = \{0,1,2\}$ and
  suppose the original policy is $d(0) = 1$, so everyone receives
  treatment level $0$. Let the harm measurement be the negative part
  function $h(y) = \max\{0, -y\}$ and the marginal laws be
  \begin{align*}
    &\P(Y(0) = 0) = 0.7,\quad \P(Y(0) = 2) = 0.3,\quad \P(Y(1) = 1) = 1, \\
    &\P(Y(2)
    = 0) = 0.4,\quad \P(Y(2) = 2) = 0.6.
  \end{align*}
  So $\mu_0 < \mu_1 < \mu_2$, but $Y(1)$ is not stochastically
  dominated by $Y(2)$. Consider the policy transition $\kappa(1 \mid
  0) = 1$, so everyone is shifted to treatment level $1$.
  Its utility and worst-case harm are given by
  \[
    \E_{\kappa}\{Y(Z')\} = \mu_1 = 1 \quad \text{and} \quad
    \sup_{\gamma \in \Gamma_{\mathrm{F}}} \EH(\kappa,\gamma) = 0.3.
  \]
  A naive approach to improve $\kappa$ is to shift the $0 \rightarrow
  1$ transitions to $0 \rightarrow 2$. Consider the policy transition
  \[
    \kappa'(0 \mid 0) = 1/3,\quad\kappa'(1 \mid 0) = 0,\quad
    \kappa'(2 \mid 0) = 2/3.
  \]
  Some $0 \rightarrow 0$ transitions are included so that the utility
  remains the same:
  \[
    \E_{\kappa'}\{Y(Z')\} = \mu_0/3 + 2\mu_2/3 = 1.
  \]
  However, it is not hard to show that
  \[
    \sup_{\gamma \in \Gamma_{\mathrm{F}}} \EH(\kappa',\gamma) =
    \frac{2}{3} \sup_{\gamma \in \Gamma_{\mathrm{F}}} \E_{\gamma}
    \{h(Y(2) - Y(0))\} = \frac{4}{3} \sup_{\gamma \in
      \Gamma_{\mathrm{F}}} \P(Y(0) = 2, Y(2) = 0) = % \frac{4}{3} \times
    % 0.3 =
    0.4.
  \]
  So $\kappa'$ has larger worst-case harm than $\kappa$. It might also
  be helpful to consider the policy transition $\kappa''(2 \mid 0) = 1$
  that shifts everyone to treatment level $2$. It is not difficult to
  verify that $\E_{\kappa''}\{Y(Z')\} = \mu_2 = 1.2$ and $\sup_{\gamma
    \in \Gamma_{\mathrm{F}}} \EH(\kappa'',\gamma) = 0.6$, so we have
  higher utility but also higher worst-case harm. To address
  this issue, a more careful modification of $\kappa$ is used in Step
  2 of the proof of \Cref{lem:leapfrogging}.
\end{example}

\section{A general lexical leapfrogging theorem}
\label{sec:leapfr-prior}

We next prove a general theorem that gives a sufficient condition for
the existence of a lexical leapfrogging solution to the counterfactual
policy optimization problem \eqref{eq:opt-eh}. Recall that in defining
the problem \eqref{eq:opt-eh}, we assume that $\Gamma = \prod_{x \in
  \mathcal{X}}\Gamma(x)$ is a product subclass of the Fr\'{e}chet
class where $\Gamma(x) \subseteq \Gamma_{\mathrm{F}}(x)$.

\begin{theorem} \label{thm:lexical-leapfrogging}
  Suppose $\Gamma = \Gamma_{\mathrm{F}}$, $h$ is subadditive, and the
  conditional distribution of $Y(z)$ given $X = x$ is
  stochastically dominated by that of $Y(z^{*}(x))$ for all $z$ and
  $x$.
    % Suppose $h$ is subadditive and $\Gamma(x)$ is non-empty and
    % closed under monotone recoupling at $(z, z^{*}(x))$ for every $z$
    % and $x$.
  Then there exists $\lambda^{*} \geq 0$ and $\alpha^{*} \in
  [0,1]$ that can be determined from $B$ such that 
  an optimal policy
  transition can be chosen as follows. For $z \neq z^{*}(x)$, let
  \begin{align*}
    \Delta(z,x) &= \E\{Y(z^{*}(x))-Y(z) \mid X=x\}, \\
    H(z,x) &= \sup_{\gamma \in \Gamma_{\mathrm{F}}(x)}
    \E_{\gamma}\{h(Y(z^{*}(x))-Y(z)) \mid X=x\},
  \end{align*}
  and define the priority score by
  \begin{equation}
    \label{eq:priority-score-general}
    \Lambda(z, x) = 
    \begin{dcases}
     \Delta(z,x)/H(z,x), & H(z,x)>0,\\
      +\infty, & H(z,x)=0<\Delta(z,x),\\
      0, & H(z,x)=\Delta(z,x)=0.
    \end{dcases}
    % \frac{\E\{Y(z^{*}(x)) - Y(z) \mid X =
    %   x\}}{\sup_{\gamma \in \Gamma_{\mathrm{F}}(x)} \E\{h(Y(z^{*}(x)) -
    %   Y(z)) \mid
    %   X = x\}}
  \end{equation}
  Set $\Lambda(z^{*}(x),x)=0$. The non optimal
  source rows $ z\neq z^{*}(x)$ of the transition are
  \begin{equation} \label{eq:lexical-leapfrogging}
    \kappa^{*}(z^{*}(x) \mid z,x)=t(z,x;\alpha^{*},\lambda^{*}),
    \qquad
    \kappa^{*}(z \mid z,x)=1-t(z,x;\alpha^{*},\lambda^{*}),
  \end{equation}
  while $\kappa^{*}(z^{*}(x) \mid z^{*}(x),x)=1$; all unlisted
  transition probabilities are zero. This transition
  solves the optimization problem \eqref{eq:opt-eh}.
\end{theorem}
\begin{proof}
  Because $\Gamma$ is a product class, equation
  \eqref{eq:sup-exp-1} still holds. By applying
  \Cref{lem:leapfrogging} to modify a policy transition $\kappa$ at
  all $x \in
  \mathcal{X}$, we arrive at a policy transition that has no less
  utility than $\kappa$, no larger worst-case harm than $\kappa$, and the leapfrogging
  structure in \eqref{eq:leapfrogging}. Thus, there exists a
  leapfrogging solution to \eqref{eq:opt-eh}.

  Next we show that the
  $\kappa^{*}$ given by \eqref{eq:lexical-leapfrogging} and
  \eqref{eq:priority-score-general} solves the original optimization
  problem \eqref{eq:opt-eh} assuming $\kappa$ is leapfrogging. With
  this assumption, we can write the objective function as
  \[
    \E_{\kappa}\{Y(Z')\} = \E\{Y(Z)\} + \E \left[ d(Z \mid X)
      \kappa(z^{*}(X) \mid Z, X) \E\{Y(z^{*}(X)) - Y(Z)
    \mid X\} \right]
  \]
  and the constraint as
  \begin{align}
    B \geq \sup_{\gamma \in \Gamma_{\mathrm{F}}} \EH_{\kappa,\gamma} =&
    \E \Big\{ \sup_{\gamma \in \Gamma_{\mathrm{F}}}
    \EH_{\kappa,\gamma}(X) \Big\} \\
    =& \E \Big[ \sup_{\gamma \in
        \Gamma_{\mathrm{F}}} \sum_{z \in \mathcal{Z}} d(z \mid X)
      \kappa(z^{*}(X) \mid z, X) \E_{\gamma}\{h(Y(z^{*}(X)) - Y(z)) \mid X\}
       \Big] \label{eq:cons-exch1} \\
    =& \E \Big[ \sum_{z \in \mathcal{Z}} d(z \mid X)
      \kappa(z^{*}(X) \mid z, X) \sup_{\gamma \in
        \Gamma_{\mathrm{F}}} \E_{\gamma}\{h(Y(z^{*}(X)) - Y(z)) \mid X\}
       \Big].
       \label{eq:cons-exch2}
  \end{align}
  The last equality is true because all the harm maximization
  problems are pairwise transport problems with respect to the same
  margin $Y(z^{*}(X))$. Therefore the pairwise optimal transport maps
  can be combined into a joint map between all the potential
  outcomes. As the objective and constraint are linear functionals of
  $\kappa$, an application of \Cref{lem:np} immediately leads to the
  optimal policy transition given by \eqref{eq:lexical-leapfrogging}
  and \eqref{eq:priority-score-general}.
\end{proof}

For binary treatment, \cref{thm:PropOptimal} is a stronger result than
\cref{thm:lexical-leapfrogging}
because it does not require any assumptions on $\Gamma$ or the harm
function $h$.

\begin{remark}\label{rmk:restricted-gamma}
For general multi-valued treatments,
\cref{thm:lexical-leapfrogging} can be extended beyond the Fr\'{e}chet
class to any $\Gamma \subseteq \Gamma_{\mathrm F}$ satisfying two
conditions. First, $\Gamma$ must be closed under monotone recoupling
(\cref{def:monotone-recoupling}) which is generally stronger than
stochastic dominance, so that \cref{lem:leapfrogging}
applies and the destination can be restricted to the optimal treatment
$z^*(x)$. Second, the constraints of $\Gamma(x)$
must be specified only in terms of the bivariate distributions between $Y(z)$ and $Y(z^{*}(x))$ given $X=x$
for each $z \in \mathcal{Z}\setminus\{z^*\}$ separately. Formally,
$\Gamma(x)$ constrains  only the first tree of a canonical vine
(C-vine) copula rooted at $z^*(x)$ \citep{aas_pair-copula_2009}, whose
edges are exactly the pairs
$\{(z, z^*(x))\}$. Because the pair-copulas of a vine may be specified freely
and independently, any combination of admissible bivariate constraints extends
to a valid joint distribution. The constraints are thus variation independent
across pairs, and the supremum over $\gamma$ may be exchanged with the
summation over the initial state~$z$ in
\eqref{eq:cons-exch1} and \eqref{eq:cons-exch2}. Moreover,
attainability of the pairwise supremum over $\Gamma$ is not required
for \cref{thm:lexical-leapfrogging}, as the priority score depends
only on the value of the supremum, not on the coupling
achieving it.
\end{remark}

The constraint in the optimization problem \eqref{eq:opt-eh}
protects against individual harm in the worst case. We now consider the
expected benefit of a transition $\kappa$,
\begin{equation*}
    \EB_{\kappa,\gamma}
      =\E_{\kappa,\gamma}\{(Y(Z')-Y(Z))_+\},
\end{equation*}
where $y_+=\max\{y,0\}$ is the positive part function. It is
interesting to ask when the lexical leapfrogging solution in
\Cref{thm:lexical-leapfrogging} also solves
\begin{equation}
  \max_\kappa\inf_{\gamma\in\Gamma_{\mathrm F}}\EB_{\kappa,\gamma}
    \qquad\text{subject to}\qquad
    \sup_{\gamma\in\Gamma_{\mathrm F}}\EH_{\kappa,\gamma}\leq B,
    \label{eq:opt-eb-eh}
\end{equation}
that is, whether that solution is Pareto efficient for maximizing
worst-case expected benefit and minimizing worst-case expected harm.

\begin{theorem}\label{thm:benefit-equivalence-total-order}
  Suppose harm is measured by the negative part function $h(y) = y_- =
  \max\{-y,0\}$. In the setting of \Cref{thm:lexical-leapfrogging}, additionally
  assume that for every $x$, all conditional potential-outcome
  marginals are totally
  ordered by stochastic dominance: $\gamma_z(x)$ either dominates or
  is dominated by $\gamma_{z'}(x)$ for all $z,z'$. Then $\kappa^*$ from
  \Cref{thm:lexical-leapfrogging} solves \eqref{eq:opt-eb-eh}.
\end{theorem}

It is essential in \Cref{thm:benefit-equivalence-total-order} that the
conditional potential-outcome marginals are totally ordered by
stochastic dominance. This is because in the absence of a total order,
it is possible to ``improve'' the policy transition $\kappa^{*}$ by
further swapping treatments that do not dominate each other; this may
decrease the average outcome and increase the worst-case harm, but it
will also increase the worst-case benefit. A counter-example can be
easily constructed even when the treatment is binary.

\section{Priority scores}
\label{sec:priority}

The optimal harm-constrained policy derived in previous sections
ranks individuals according to a priority score involving the supremum
\[
  \sup_{\gamma \in \Gamma} \E\{h(Y(z^{*}(x)) - Y(z)) \mid X = x\},
\]
where $\Gamma$ encodes assumptions on the dependence between
potential outcomes. This supremum is not always easy to compute, and
in this section we derive closed-form expressions for the supremum and
priority score for certain copula models of binary and general
real-valued outcomes. To simplify the exposition, we will omit
the covariates $X$ and consider a baseline treatment $z$ and optimal
treatment $z^*$.

\subsection{Binary Outcomes}
\label{sec:priority-binary}
Since $Y$ is binary, the argument of $h$ takes values in
$\{-1, 0, 1\}$, so any harm function satisfying \cref{def:harm-func}
reduces to $h(y) = c \cdot \mathbf{1}_{\{y < 0\}}$ for some $c > 0$. Setting
$c = 1$ without loss of generality, we are left to determine the supremum over $\gamma \in \Gamma$ of the probability of harm
\begin{equation}\label{eq:ph}
    \PH=\P(Y(z^*)<Y(z)).
\end{equation}
Moreover, the joint distribution of $Y(z), Y(z^*)$ can be uniquely
determined by the marginals $\gamma_z, \gamma_{z^*}$, and the odds
ratio
\begin{equation}
    \OR(z,z^*) = \frac{\P\{Y(z) = 1, Y(z^*) = 1 \} / \P\{Y(z) =0, Y(z^*)
      =1 \}}{\P\{Y(z) =1, Y(z^*) =0  \} / \P\{Y(z) = 0, Y(z^*)
      = 0\}}.
\end{equation}
The Fr\'{e}chet class puts no constraint on this odds ratio (it is between
$0$ and $\infty$, inclusive on both ends). In practice, one may want to put constraints on the dependence between $Y(z)$ and $Y(z^*)$. Let us therefore consider the following class
\[
  \Gamma(\theta_L) = \{\gamma \in
  \Gamma_{\text{F}}: \OR(z,z^*) \geq \theta_L \, \forall \, z\in \mathcal{Z} \},
\]
where $\theta_L:\mathcal{Z} \to [0, \infty]$, so taking $\theta_L = 0$
corresponds to using the entire Fr\'{e}chet class, taking $\theta_L =
1$ corresponds to assuming the potential outcomes have a non-negative
correlation, and taking $\theta_L = \infty$ corresponds to assuming
comonotone potential outcomes (perfect positive
dependence). \citet{wu_safe_2025} instead parametrize the bivariate
dependence by Pearson's correlation, which is more difficult to interpret
because, unlike the odds ratio, Pearson's correlation is not variation
independent of the marginals.

Closed-form formulas for the priority scores under the class
$\Gamma(\theta_L)$ can be found in \cref{prop:quadratic} in the
\cref{app:proofs-priority}. Here we give the formula in two special
cases:
\begin{align*}
  \Lambda_0(z) &= \frac{\E\{Y(z^{*}) - Y(z)\}}{\min\{\gamma_z, 1 - \gamma_{z^*}\}},&&
    \text{when}~\theta_L(z) \to  0~\text{(Fr\'{e}chet class)}, \\
  \Lambda_1(z) &=  \frac{\E\{Y(z^{*}) - Y(z)\}}{\gamma_z \cdot \{1 - \gamma_{z^*}\}},&&
    \text{when}~\theta_L(z) = 1~\text{(Non-negative correlation)},
\end{align*}
where $\Lambda(z)$ is sub-indexed by $\theta_L(z)$.

\subsection{General Outcomes}

When the outcome $Y$ is a general real-valued variable, many different
choices of the harm function $h$ are possible, but the supremum
\[
  \sup_{\gamma \in \Gamma_{\mathrm{F}}} \E\{h(Y(z^*) - Y(z)) \}
\]
generally requires numerical approximation. Closed-form formulas can
be obtained for specific harm functions using results from the optimal
transport literature. More concretely, if the function $f(x,y) :=
h(x-y)$ is sub-modular on $\R^2$, i.e.\ $f(x,y) + f(x',y') \leq
f(x,y')+f(x',y)$ for all $x \leq x'$ and $y\leq y'$ in $\R$, then the
supremum is achieved at the counter-monotone coupling of $Y(z^*)$ and
$Y(z)$ \citep{tchen_inequalities_1980}. However, only a nonnegative scalar multiple of the negative
part $h(y)=y_-$ fulfils sub-additivity and sub-modularity; see
Lemma~\ref{lem:neg-part} in \cref{app:proofs-priority}. For this
reason, we will focus on $h(y) =
y_-$ hereafter.
It follows from \citet[Cor.~2.3]{tchen_inequalities_1980} or
\citet[Thm.~2]{cambanis_inequalities_1976} that\footnote{Note that
  \citeauthor{tchen_inequalities_1980} refers to a sub-modular
  function as ``sub-additive'' and
  \citeauthor{cambanis_inequalities_1976} as ``quasi-antitone''.}
\begin{equation} \label{eq:harm-countermonotonic}
  \sup_{\gamma \in \Gamma_{\mathrm{F}}} \E[\{Y(z^*) - Y(z)\}_-]
  = \int_0^1 \{\gamma_{z^*}^{-1}(u) - \gamma_z^{-1}(1-u)\}_-\,\mathrm{d}u,
\end{equation}
where $\gamma_{z}^{-1}(u) := \inf\{t \colon \gamma_{z}(t) \geq u\}$ is
the quantile function of $Y(z)$, $z \in \mathcal{Z}$.
% \begin{remark}
%     If the potential outcomes are binary, the proposition above recovers the worst case expected harm under no constraint on the odds ratio ($\theta_L \to 0$) derived in Proposition~\ref{prop:quadratic}.
%   \end{remark}

Closed-form expressions can also be obtained beyond the agnostic
Fr\'{e}chet class~$\Gamma_{\mathrm{F}}$ (e.g., using odds ratios in the
binary setting). For continuous outcomes, we will
consider multivariate Gaussian copulas, allowing singular
positive-semidefinite correlation matrices:
\begin{equation*}
    \Gamma(\rho_L) =\Big\{\gamma~\text{has a Gaussian copula with
      correlation $R$} \colon R_{z,z^*} \geq \rho_L,\, \forall z \in \mathcal{Z}
    \Big\},~\rho_L \in [-1,1];
\end{equation*}
the distribution function $\gamma$ of $Y(0),\ldots,Y(M)$ has a Gaussian copula if
\begin{equation*}
    \gamma(y)=\P\big(Y(0)\leq y_0, \ldots, Y(M)\leq y_{M}\big) =
    \Phi_R\Big(\Phi^{-1}(\gamma_0(y_0)),\ldots,\,\Phi^{-1}(\gamma_{M}(y_{M}))\Big),
\end{equation*}
where $\Phi$ is the distribution function of a standard normal
random variable and $\Phi_R$ is the cdf of an $(M+1)$-dimensional Gaussian
with mean zero, unit variance and correlation matrix $R$.
Unlike the binary case, imposing a Gaussian copula structure is a
substantive restriction on the form that the joint distribution $\gamma$
can take: $\Gamma(\rho_L)$ is a proper subset of
$\Gamma_{\mathrm{F}}$. Nonetheless, the countermonotonic coupling is
recovered by taking $\rho_L = -1$. The next result further shows that
the priority score is indeed continuous at $\rho_L = -1$.

\begin{prop}\label{prop:copula} Assume that the marginal distribution functions $\gamma_z$ are continuous and let $\rho_L \in (-1,1)$. Then,
    \begin{align*}
        &\sup_{\gamma \in \Gamma(\rho_L)} \E[\{Y(z^*) -
      Y(z)\}_- ]\\
      =& \int_{-\infty}^\infty \int_{0}^\infty \Phi\left(\frac{\rho_L\, \Phi^{-1}(\gamma_{z^*}(y)) - \Phi^{-1}(\gamma_z(y+t))}{\sqrt{1-\rho_L^2}}\right)\, \mathrm{d}t\, \mathrm{d}\gamma_{z^*}(y),
    \end{align*}
    and this converges to \eqref{eq:harm-countermonotonic} when
    $\rho_L \to -1$.
\end{prop}

Under the additional assumption that $Y(z)$ is stochastically
dominated by $Y(z^{*})$, \cref{thm:lexical-leapfrogging} shows that
the bound in \Cref{prop:copula} yields a harm-aware priority score. In
the special case of multivariate Gaussian outcomes, the priority score
has a closed-form expression, depending on $\rho_L$. However, when the same $\rho_L$ is used across the pairs
being compared, its ordering is equivalent to ordering by the benefit of switching to the conditionally optimal arm normalized by the conditional standard deviation.

\begin{prop}\label{prop:gaussian}
  Let $\rho_L<1$. Suppose $(Y(z): z \in \mathcal{Z})$ given $X=x$ follows a
  multivariate Gaussian distribution with positive conditional variances and pairwise correlation lower
  bounded by $\rho_L$ (so the distribution belongs to
  $\Gamma(\rho_L)$). Write $\mu_z(X)=\E\{Y(z)\mid X\}$ and
  $\sigma^2_z(X)=\Var\{Y(z)\mid X\}$. If $Y(z)$
  is stochastically dominated by $Y(z^{*})$ for every $z \in
  \mathcal{Z}$, then
\begin{enumerate}
  \item[(i)] the arms share a common dispersion: $\sigma_z(x) =
    \sigma(x)$ for all $z \in \mathcal{Z}$;
  \item[(ii)] the priority score in \cref{thm:lexical-leapfrogging} is
    given by
    \[
      \Lambda_{\rho_L}(z,x) =
      g\Big(\frac{r(z,x)}{\sqrt{2(1-\rho_L)}}\Big), \quad r(z,x) =
      \frac{\mu_{z^*}(x) - \mu_z(x)}{\sigma(x)},
    \]
    where $g(r) =  r/\{\varphi(r) - r\,\Phi(-r)\}$ is an increasing
    function.
\end{enumerate}
\end{prop}

% \Cref{tab:cont-example} provides a simple demonstration with
% two treatment arms and three covariate groups.

% \begin{table}[h]
% \caption{Three groups illustration of \cref{prop:gaussian}. $R$ is the ranking
% position under the CATE $\tau(x)$ and under the benefit-to-dispersion ratio
% $\tau(x)/\sigma(x)$, which ranks every class $\Gamma(\rho_L)$ identically. Bold entries differ from the CATE ranking.}
% \label{tab:cont-example}
% \centering
% \begin{tabular}{c|c|rr|rr}
% \toprule
% & & \multicolumn{2}{c|}{CATE} & \multicolumn{2}{c}{Harm-aware score}\\
% $x$ & $\sigma(x)$ & $\tau(x)$ & $R$ & $\tau(x)/\sigma(x)$ & $R$ \\
% % & $\Lambda_{\rho_L}(x)$ \\
% \midrule
% $G_1$ & $0.5$ & $1.0$ & $2$ & $2.00$ & $\mathbf{1}$ \\ % & $12.00$ \\
% $G_2$ & $1$   & $1.0$ & $2$ & $1.00$ & $2$ \\ % & $2.53$  \\
% $G_3$ & $2$   & $1.3$ & $1$ & $0.65$ & $\mathbf{3}$  \\ % & $1.26$  \\
% \bottomrule
% \end{tabular}
% \end{table}
The priority scores obtained in this section depend on the joint
distribution of the potential outcomes only through the marginal
distribution of $Y(z), z\in \mathcal{Z}$ given $X$ and the sensitivity
parameter $\theta_L$ or $\rho_L$. Plug-in estimators can be easily
obtained by substituting an estimate of those distributions, for
instance from quantile regression \citep{koenker_handbook_2017}. This
is further simplified in the multivariate Gaussian case, as only the
conditional mean and variance functions are required.

\section{Numerical experiments}\label{sec:numerical}
Next we illustrate the finite-sample behavior of the priority scores using a real dataset from the ongoing
I-SPY2 breast-cancer platform trial \citep{barker_ispy2_2009,wang_yee_ispy2_2019}.
I-SPY2 is a multi-arm, Bayesian response-adaptively randomized platform trial in the neoadjuvant setting. We draw on the
patient-level annotation from \citet[Table~S2]{wolf_redefining_2022},
reproduced in \cref{tab:ispy2_data} here, comprising 987
patients distributed across $10$ arms, including a shared
control arm. We use pathologic complete response (pCR) as the binary
outcome $Y$ and stratify the patients by the standard clinical
receptor-based subtype (HR/HER2). This analysis is for the sole purpose of illustrating different priority scores developed above and should not be used as clinical advice.
Since arm availability and response-adaptive randomization depended on finer biomarker strata and trial history, the crude (arm, subtype) response rates need not identify the marginal distribution of potential outcome.

\begin{table}[tbp]
  \centering
  \caption{I-SPY2 cells by receptor subtype and treatment arm. Cell entries are
  $n_{\mathrm{pCR}}/n_{\mathrm{cell}}$ with the raw pCR rate in
  parentheses; -- marks combinations with no patients in the
  data. The pooled column aggregates across receptor subtypes; bold
  entries indicate the highest pCR rate within each column. Full names
  of the treatment arms can be found in
  \citet[Table~S2]{wolf_redefining_2022}.}
  \label{tab:ispy2_data}
  \begin{tabular}{lccccc}
\toprule
Arm & HR-HER2+ & HR+HER2- & HR+HER2+ & TN & Pooled \\
\midrule
AMG386 & 2/4\,(0.50) & 10/62\,(0.16) & 5/15\,(0.33) & 23/53\,(0.43) & 40/134\,(0.30) \\
Ctr & 5/12\,(0.42) & 14/94\,(0.15) & 3/19\,(0.16) & 16/85\,(0.19) & 38/210\,(0.18) \\
Ganetespib & -- & 6/48\,(0.12) & -- & 19/45\,(0.42) & 25/93\,(0.27) \\
Ganitumab  & -- & 8/58\,(0.14) & -- & 16/48\,(0.33) & 24/106\,(0.23) \\
MK2206 & 12/18\,(0.67) & 5/28\,(0.18) & 5/16\,(0.31) & 13/32\,(0.41) & 35/94\,(0.37) \\
N  & 13/23\,(0.57) & 3/17\,(0.18) & 13/42\,(0.31) & 12/32\,(0.38) & 41/114\,(0.36) \\
Pembro  & -- & \textbf{12/40\,(0.30)} & -- & \textbf{19/29\,(0.66)} & 31/69\,(0.45) \\
Pertuzumab  & \textbf{12/15\,(0.80)} & -- & 14/29\,(0.48) & -- & \textbf{26/44\,(0.59)} \\
TDM1/P & 12/17\,(0.71) & -- & \textbf{18/35\,(0.51)} & -- & 30/52\,(0.58) \\
VC & -- & 5/32\,(0.16) & -- & 22/39\,(0.56) & 27/71\,(0.38) \\
\midrule
Total ($n$) & 89 & 379 & 156 & 363 & 987 \\
Pooled rate & 0.63 & 0.17 & 0.37 & 0.39 & 0.32 \\
\bottomrule
\end{tabular}
\end{table}

\subsection{Empirical simulation studies}\label{sec:sim}

The harm-aware priority scores for the treatment-subtype combinations
can be easily obtained by plugging the empirical probabilities into
the formulas in \Cref{sec:priority-binary}. However, given the modest
sample size in \cref{tab:ispy2_data}, we first use empirical
simulation studies to assess the uncertainty in the ranking obtained
from those priority scores.

\begin{figure}[t]
    \centering
    \includegraphics[width=\linewidth]{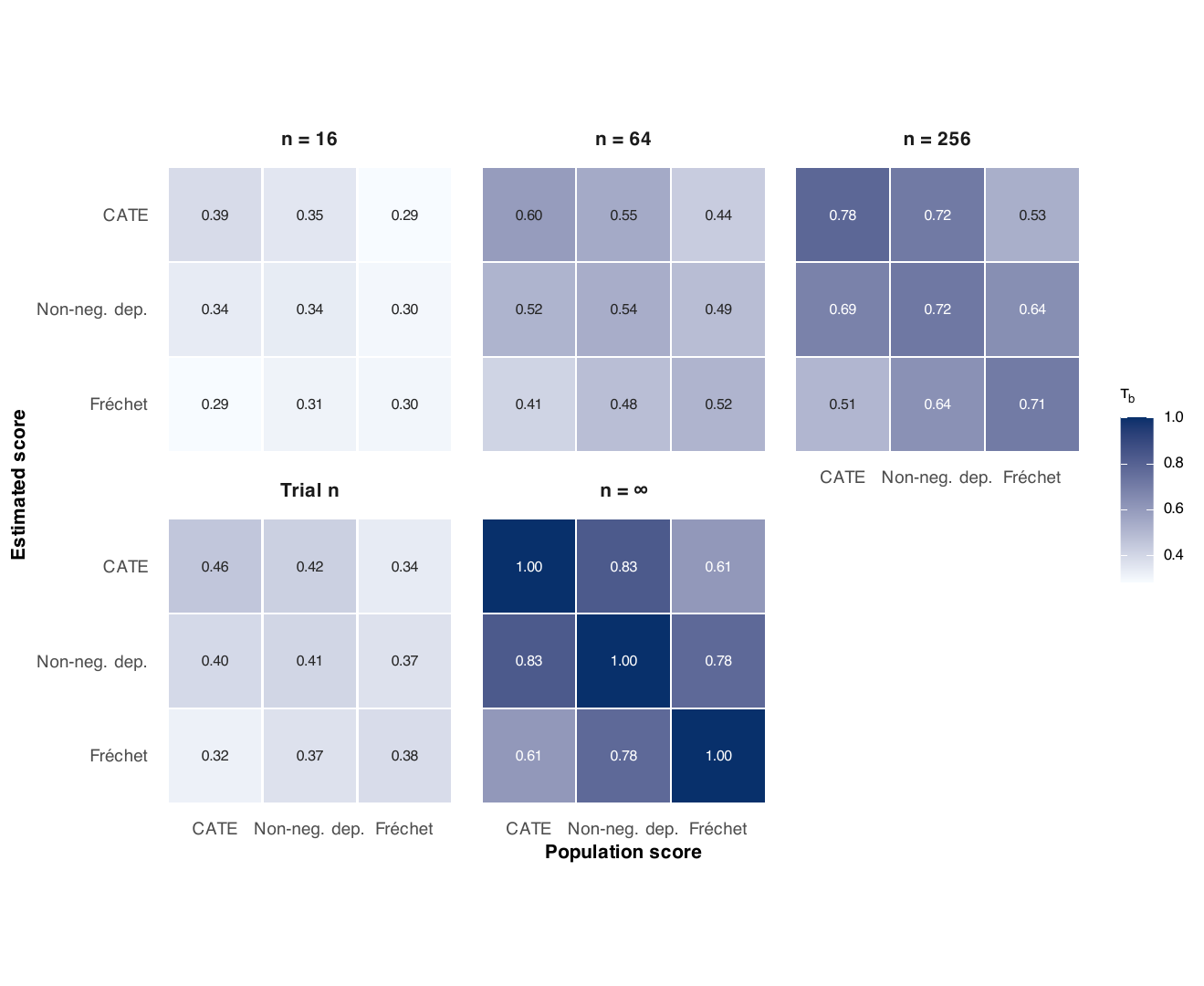}
    \caption{Median Kendall's $\tau_b$ between each estimated ordering (rows)
    and each population ordering (columns), at different sample sizes $n$. Off-diagonal entries measure
    disagreement between methods; diagonal entries show the
    median rank correlation with the corresponding population ranking.}
    \label{fig:heatmap}
\end{figure}

\begin{figure}[t]
  \centering
  \includegraphics[width=0.9\linewidth]{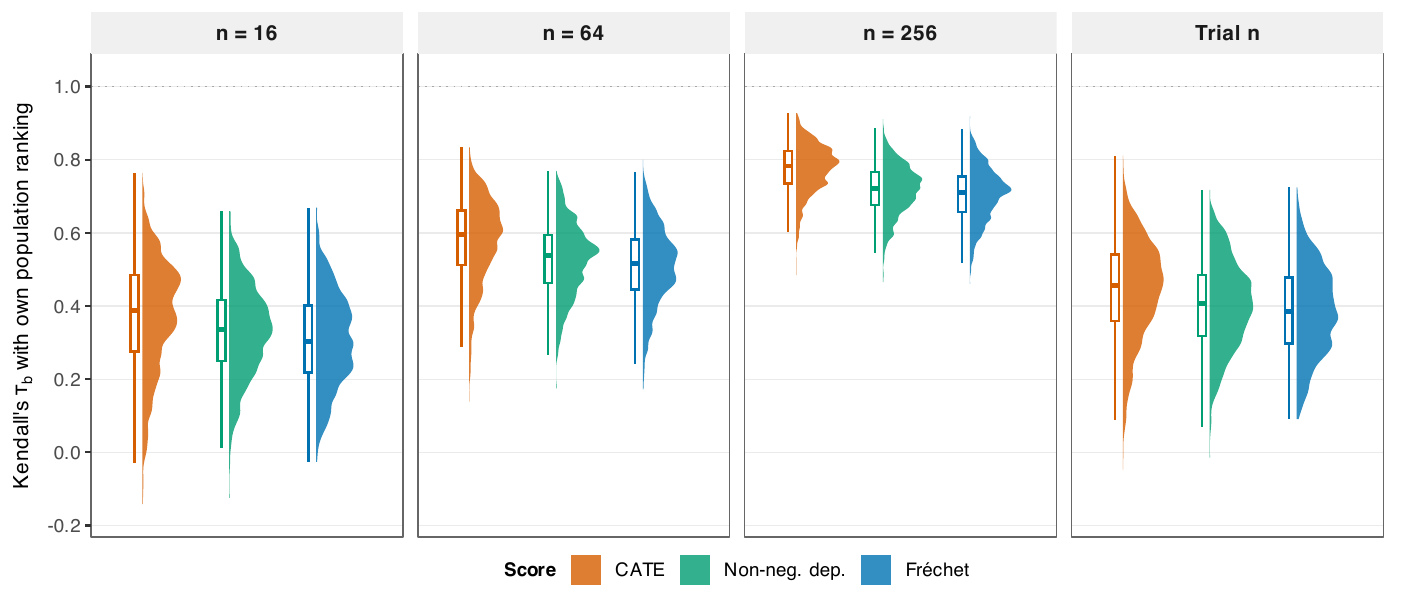}
  \caption{Recovery of each score's own population ranking: per-replicate
    Kendall's $\tau_b$ between the estimated and population ranking, by method,
    across a common-$n$ grid and the trial-$n$ regime.}
  \label{fig:recovery}
\end{figure}

\begin{figure}[t]
  \centering
  \includegraphics[width=0.9\linewidth]{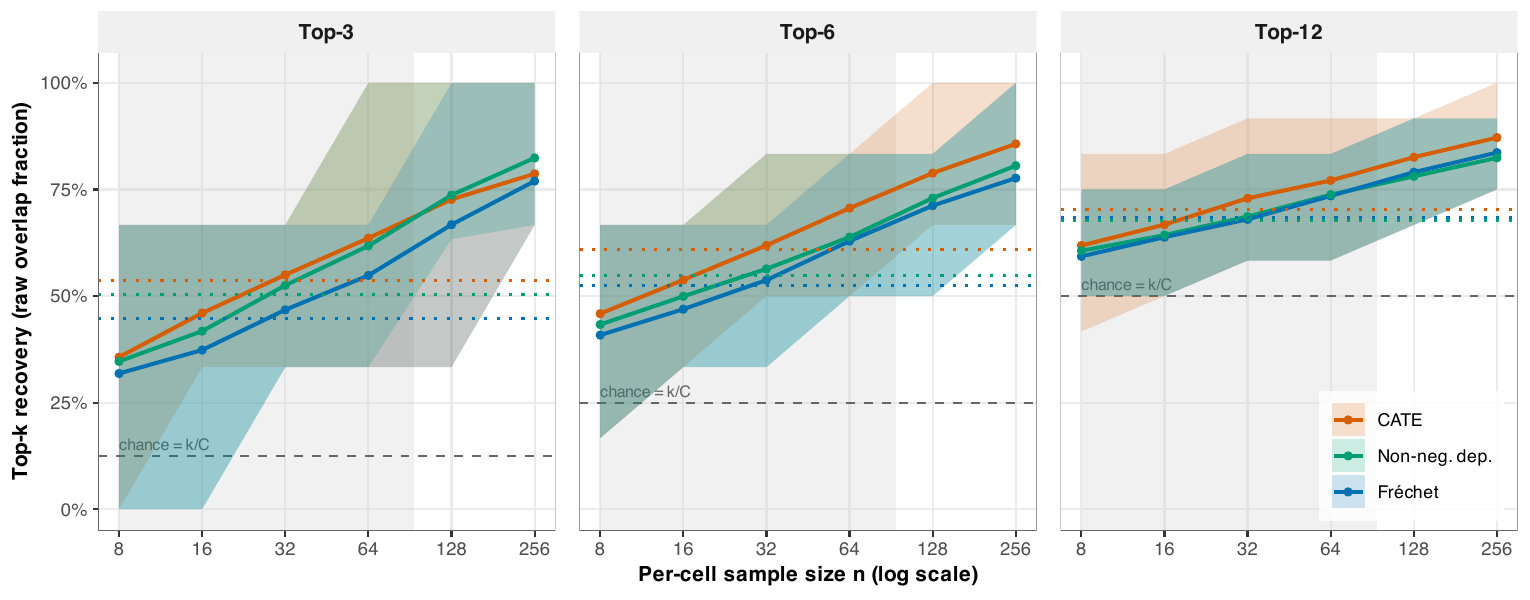}
  \caption{Top-$k$ recovery: mean overlap fraction between the estimated and
    population top-$k$ priority sets against per-cell size $n$ (log scale). The dashed horizontal line is the expected overlap for two independent uniformly random $k$-subsets for the $C=24$ candidate switches; while the dotted horizontal lines give the mean recovery for each method under the trial-$n$ regime.  
    The colored band gives the 10\% and
    90\% quantiles of the recovery fraction across replications; the grey vertical band marks the range of observed cell sizes.
    %Ties at the $k$-boundary are broken at random in the estimated sets. 
     }
  \label{fig:topk}
\end{figure}

To this end, we adopt flat priors on the cell-specific response rates, so
that given $y_z(x)$ responders out of $n_z(x)$ patients the posterior
distribution is $\mathrm{Beta}(y_z(x)+1,\,n_z(x)-y_z(x)+1)$. To create
the empirical simulation, we then fix the true response rate at the
corresponding posterior mean $\{y_z(x)+1\}/\{n_z(x)+2\}$ and generate Binomial
data for each cell with different sample sizes. We then re-estimate the
cell-specific response rates and recompute the priority scores and
rankings. \Cref{fig:heatmap,fig:recovery,fig:topk} report 1000 replications under a common per-cell size $n$ and a \emph{trial-$n$} regime that draws each cell at its own observed size.

\Cref{fig:heatmap} shows substantial differences among the rankings
obtained by CATE, the non-negative dependence score $\Lambda_1$, and
the Fr\'{e}chet score $\Lambda_0$. In the population ($n = \infty$),
the CATE score has a $0.83$ rank correlation (Kendall's $\tau_b$) with $\Lambda_1$ and $0.61$ rank correlation with the
Fr\'{e}chet score. Not surprisingly, as the sample
size increases, the finite-sample rankings show higher correlation
with the corresponding population rankings (diagonal entries in
\Cref{fig:heatmap}). \Cref{fig:recovery} further shows that the
ranking correlation can have large variability for modest sample
sizes. \Cref{fig:topk} shows the recovery of top-$k$ priority
treatment-subtype groups ($k\in\{3,6,12\}$) with finite sample
sizes. With the sample size in the actual data, about half of the
top-$3$ or top-$6$ groups are expected to be recovered.

\subsection{Analysis of the I-SPY2 trial}\label{sec:real}

\begin{figure}[t]
  \centering
  \includegraphics[width=\linewidth]{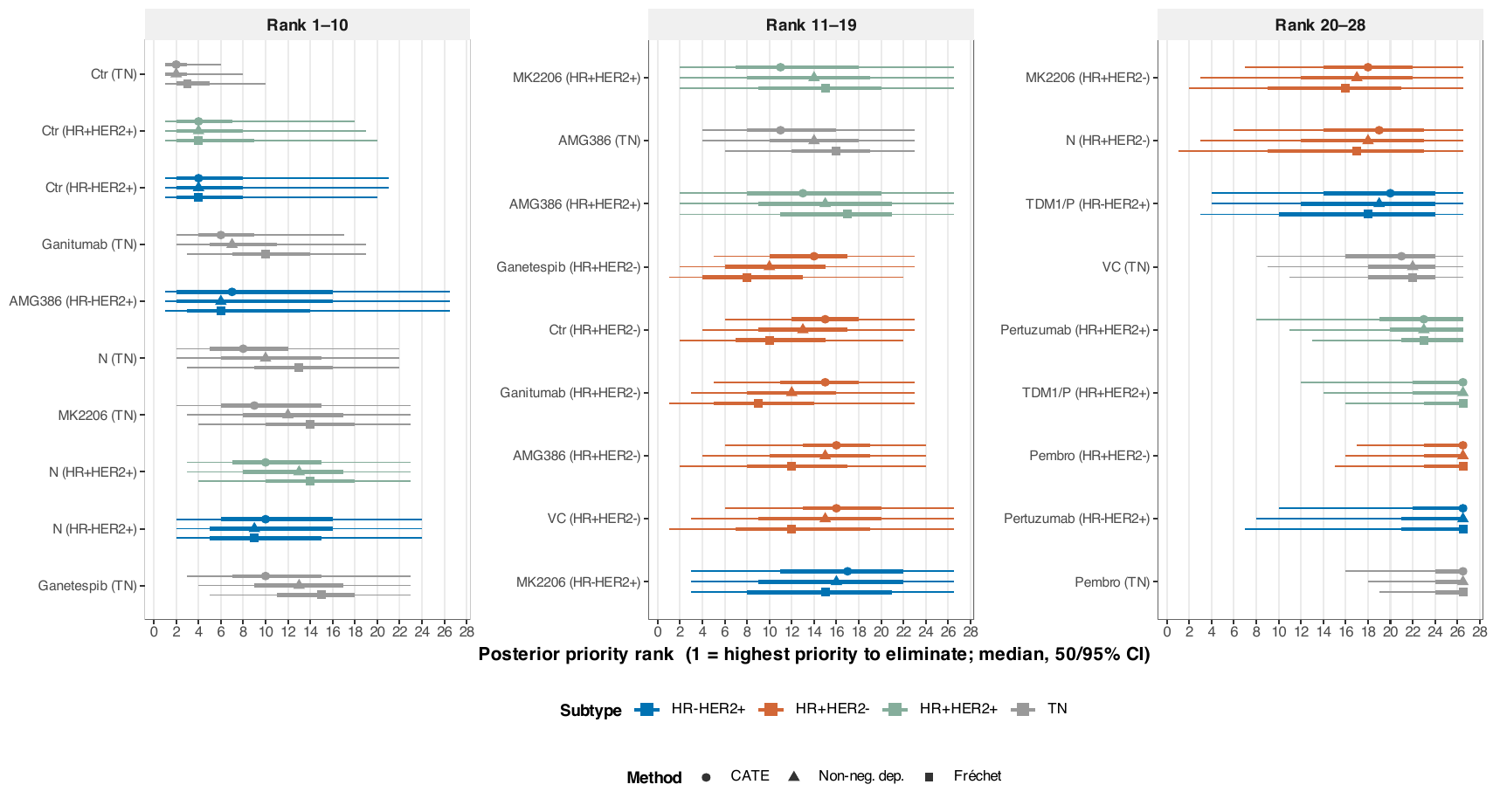}
  \caption{Posterior priority-rank credible intervals (median, 50\% and 95\%) for
    each candidate under CATE, non-negative dependence and Fr\'{e}chet scores, ordered by CATE median
    rank. Rank~$1$ is highest priority. For each subtype, the arm most often having the highest response probability across posterior draws appears at the bottom of the ordering. }
  \label{fig:ranks}
\end{figure}

\begin{figure}[t]
  \centering
  \includegraphics[width=\linewidth]{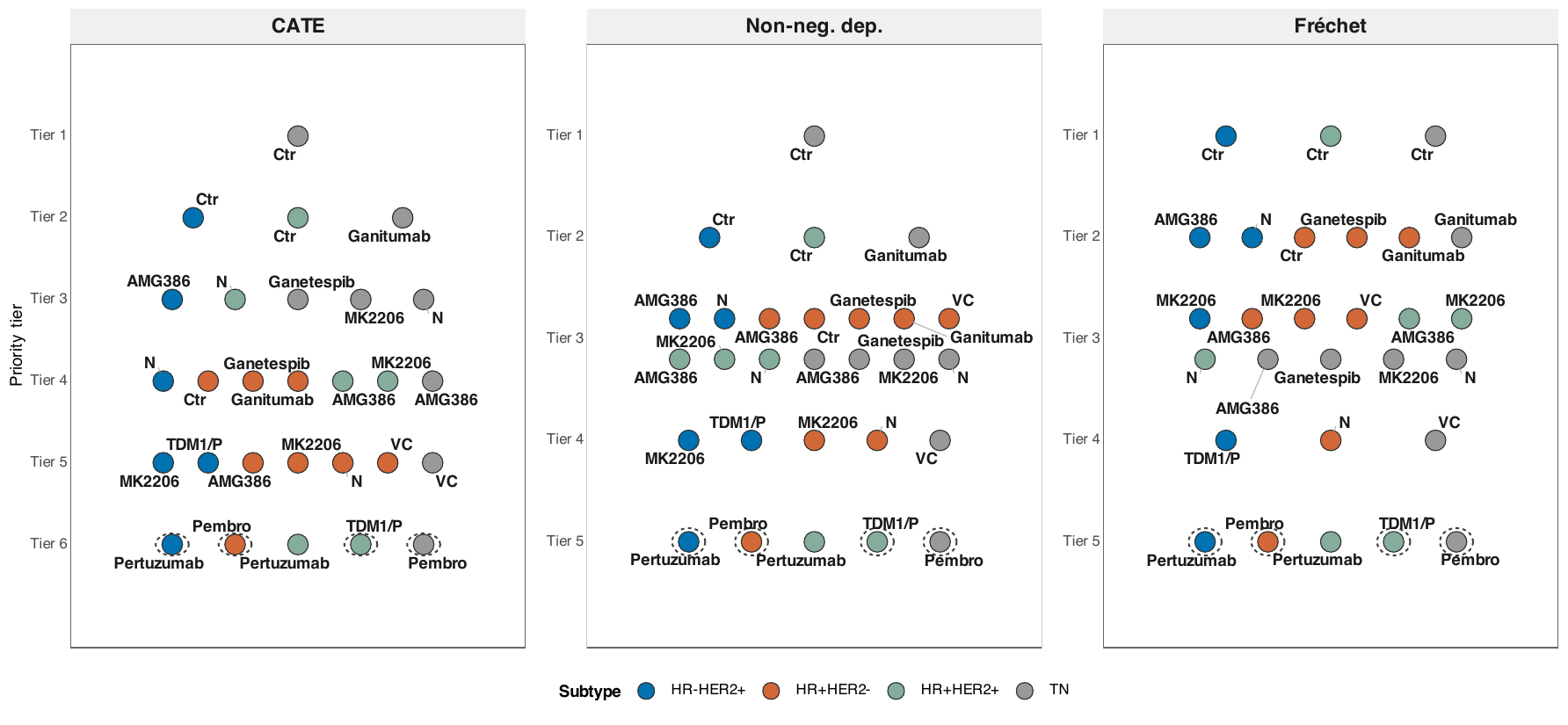}
  \caption{Exploratory posterior priority tiers at $\alpha=0.30$ for CATE, non-negative dependence and
    Fr\'{e}chet scores. Nodes are (subtype, arm) candidates colored by subtype; Tier~1 at
    the top is the highest priority to eliminate. A dashed border marks an arm optimal within its subtype with posterior probability above 0.5.}
  \label{fig:hasse}
\end{figure}

\Cref{fig:ranks} shows the posterior credible intervals for the
treatment-subtype groups, based on 10,000 independent draws from the cell-specific Beta posterior distributions. The
credible intervals show substantial uncertainty in the individual
rankings, which is consistent with the empirical simulation in
\Cref{sec:sim}.

For this reason, we consider another approach arranging the treatment-subtype
groups into priority ``tiers''. We say a group $(z,x)$ dominates another
group $(z',x')$ if $\Lambda(z,x) > \Lambda(z',x')$ with posterior
probability larger than $1 - \alpha$. This pairwise relation can be non transitive and cyclic, so it is not a partial order in general, though no cycles arise for the analyzed data.

We can then organize different
$(z,x)$ into tiers, starting from the
bottom up: the last tier contains those that do not dominate any other
group, the second-to-last tier contains those that dominate only groups
in the last tier, and so on. We will refer the last and
highest-priority tier as tier~1. \Cref{fig:hasse} shows the results
obtained from CATE, $\Lambda_1$ (non-negative dependence), and
$\Lambda_0$ (Fr\'{e}chet) with $\alpha = 0.3$. The three scores give
substantially different priority tiers. For example, both CATE and
$\Lambda_1$ have Ctr/TN as their top priority, while the Fr\'{e}chet score has three groups---Ctr/HR-HER2+, Ctr/HR+HER2+, Ctr/TN---in its top
priority tier.

\section{Discussion}\label{sec:discussion}

In practice, posterior rankings such as those in
\Cref{fig:ranks} and the priority tiers in \Cref{fig:hasse} could
inform decisions about which treatment arms to discontinue within a
clinical subtype. For example, subject to appropriate clinical review,
regulators might decide based on these results to discontinue certain
treatment--subtype combinations in the two
highest-priority tiers under the Fr\'{e}chet score, while retaining
the rest of the options for doctors and patients to choose from when
additional clinical information is provided. This regulatory process
can also be repeated as further data accumulate or new treatments
become available.

Counterfactual harm therefore provides a perspective distinct from
standard precision medicine and policy learning. By acknowledging
that a policy transition can leave some individuals worse off and by
placing an explicit bound on such losses, our framework generates a
lexical solution path toward the optimal policy. Lexical solutions can
also arise when policy optimization is subject to external
resource constraints; see \Cref{sec:revi-policy-optim}. Here in our
framework, however, the constrained quantity is not an exogenous
financial or capacity cost but an intrinsic, causal, moral loss
induced by the policy transition. In this vein, perhaps the most
interesting finding is that the stochastic-dominance
and recoupling conditions of \Cref{lem:leapfrogging} imply a
``leapfrogging'' property: an optimal transition moves directly to the
conditionally best arm rather than proceeding through intermediate
arms. Ordinary resource constraints do not imply this property, since
an otherwise attractive transition may be deferred or rejected simply
because it is prohibitively costly.

Lexical priority has long been studied in social choice, welfare
economics, and political philosophy since the monumental work of
\citet{rawls1971theory}. However, we could not find any existing
lexical social orderings in the literature with the leapfrogging
structure obtained here. Rawls himself used maximin reasoning behind the veil of
ignorance to motivate the difference principle, which evaluates social
and economic inequalities by their effects on the prospects of the least
advantaged. Strict maximin has been criticized for giving absolute priority to an
arbitrarily small improvement to the worst-off over
potentially large losses to others
\citep{arrow1973ordinalist,harsanyi1975maximin}. Subsequent approaches
such as prioritarianism give additional but generally finite weight to
benefits accruing to worse-off individuals rather than assigning them
absolute lexical priority
\citep{parfit1997equality,adler2012wellbeing}.

Our framework thereby occupies another middle ground between
unconstrained aggregate-welfare maximization and strict lexical
priority. It retains the consequentialist objective of maximizing
expected population benefit, but restricts the feasible policies
through a bound on worst-case counterfactual harm. In other words, it
combines an aggregative welfare objective with an ethically motivated
constraint, rather than adopting either pure utilitarianism or
Rawlsian maximin. What is interesting is that we recover Rawlsian
lexical priority (under appropriate conditions) but not the maximin
structure; in fact, the leapfrogging structure is central in
establishing lexical ordering in the proof of
\Cref{thm:lexical-leapfrogging}.

Our perspective is transition-based rather than purely
distributive. From the policymaker’s perspective, we seek policy
transitions that improve expected population outcomes without
exceeding a bound on counterfactual harm. The relevant individuals are
therefore not necessarily those who are worst off in absolute terms—as
measured by welfare, wealth, resources, capabilities, or primary
goods—but those who would be made worse off by the contemplated
transition. Counterfactual harm is a within-person, cross-policy
comparison, whereas much of distributive justice evaluates
individuals’ positions within the resulting allocation. These
perspectives are complementary: one concerns the distribution of
social advantage, while the other concerns who bears the losses
created by social change.

Related distributive concerns have entered policy learning through two
main routes. One modifies the objective, replacing average welfare
with an equality-minded or distributional welfare functional
\citep{kitagawa_equality-minded_2021,cui_policy_2025}; the other
imposes a harm constraint \citep{wu2026individualrisk}, as we do here.
The asymmetric counterfactual utilities considered in
\citet{benmichaelpolicy2024} also belong to the second route by
interpreting a penalized formulation as the Lagrangian relaxation of
our constrained problem.
Several related approaches move beyond unconstrained mean-welfare
maximization. \citet{viviano_fair_2024} select the
fairest allocation on the welfare-efficient frontier under a non-maleficence
principle, and \citet{fang_fairness-oriented_2023} require quantiles of the outcome
distribution to exceed a threshold. These criteria concern marginal or
group-conditional policy outcomes and do not require the coupling of
potential outcomes that determines our counterfactual harm.

Our framework has several limitations.
Our analysis treats benefit and counterfactual harm as changes in the
same scalar outcome. Applications with distinct or multidimensional
safety outcomes would require defining the relevant counterfactual
safety comparisons and specifying an appropriate utility function and
joint model for the multivariate potential outcomes. Moreover, our
scalar priority ordering relies on a single linear harm
constraint. Resource, fairness,
or subgroup-specific harm constrains can be added computationally,
but generally produce a multi-constraint optimization problem without
the present one-score threshold solution. Finally, because the harm
budget controls population-level worst-case expected loss, it does not
guarantee that every individual or subgroup is protected from harm.

\appendix
\section{Proofs}\label{app:proofs}
\subsection{\texorpdfstring{Proof of Results in \cref{sec:optimization-bin}}{Proof of Results in ~\ref{sec:optimization-bin}}}\label{app:proofs-bin}
\begin{proof}[Proof of Theorem~\ref{thm:PropOptimal}]
  Consider any policy transition $\kappa$ such that $\kappa(0 \mid 1,
  x) > 0$ for some $x \in \mathcal{X}$ such that $\tau(x) \geq 0$. Then
  consider the policy transition $\kappa'$ that only differs from
  $\kappa$ in that
  \[
    \kappa'(0 \mid 1, x) = 0,\quad\kappa'(1 \mid 1, x) = \kappa(1 \mid
    1, x) + \kappa(0 \mid 1, x).
  \]
  It is not difficult to see that $\E_{\kappa'} \{Y(Z')\} \geq
  \E_{\kappa} \{Y(Z')\}$ (because more individuals with covariates $x$
  receive the better overall treatment) and $\EH_{\kappa',\gamma} \leq
  \EH_{\kappa,\gamma}$ for all $\gamma \in \Gamma$ (because fewer
  transitions are made). Similarly, one can remove all transitions
  from $0$ to $1$ if $\tau(x) \leq 0$. So it suffices to only consider
  policy transitions $\kappa$ that satisfy $\kappa(0 \mid 1, x) = 0$ if
  $\tau(x) \geq 0$ and $\kappa(1 \mid 0, x) = 0$ if $\tau(x) \leq
  0$. With this assumption, the optimization problem \eqref{eq:opt-eh}
  can be rewritten as the continuous knapsack problem
  \eqref{eq:np-knapsack} with
  \begin{align*}
    a(X) &= \mathds{1}_{\{\tau(X) > 0\}} \cdot (1 - d(X)) \tau(X) -
           \mathds{1}_{\{\tau(X) < 0\}} \cdot d(X)
           \tau(X), \\
    b(X) &= \mathds{1}_{\{\tau(X)
           > 0\}} \cdot (1-d(X)) \sup_{\gamma \in \Gamma}
           \E_{\gamma(X)}\{h(Y(1) - Y(0)) \mid X\} \\
    &+
           \mathds{1}_{\{\tau(X) < 0\}} \cdot d(X) \sup_{\gamma \in \Gamma(X)} E_{\gamma}\{h(Y(0) - Y(1)) \mid
           X\}, \\
    c(X) &= \mathds{1}_{\{\tau(X) > 0\}} \cdot \kappa(1 \mid 0, X)  +
           \mathds{1}_{\{\tau(X) < 0\}} \cdot \kappa(0 \mid 1, X) .
  \end{align*}
  The conclusion then follows from applying \Cref{lem:np}.
\end{proof}

\subsection{\texorpdfstring{Proof of Results in \cref{sec:multi-valu-treatm}}{Proof of Results in ~\ref{sec:multi-valu-treatm}}}\label{app:proofs-multi}
\begin{proof}[Proof of \Cref{lem:leapfrogging}]
  As mentioned in the main text, the proof consists of three
  consecutive transformations of $\kappa$ that never decrease the mean
  potential outcome or increase the worst-case counterfactual
  harm. Without loss of generality, assume $\mu_0 \leq \mu_1 \leq
  \dots \leq \mu_M$ and $z < z^{*}$ (as $\gamma_{z}$ is
  stochastically dominated by $\gamma_{z^{*}}$).

  \noindent \textbf{Step 1:} We first remove all ``downward
  transitions'' in $\kappa$. That is, suppose
  $\kappa_1$ is the policy transition that only differs from $\kappa$ in
  the following entries:
  \[
    \kappa_1(z_2 \mid z_2) = \kappa(z_2 \mid z_2) + \sum_{z_1<z_2}
    \kappa(z_1 \mid z_2)
    \quad \text{and} \quad \kappa_1(z_1 \mid z_2) = 0,~\text{for
      all}~z_1 < z_2.
  \]
  It is obvious that $\kappa_1$ satisfies the first two properties of
  $\kappa^{*}$ in the Lemma statement.

  \noindent\textbf{Step 2:} We next remove ``upward transitions'' into
  \emph{or} out of $z$ in $\kappa_1$.
  As long as there exists $0 \leq \underline{z} < z < \overline{z}
  \leq M$ such that $\kappa_1(z \mid
  \underline{z}) > 0 $ and $\kappa_1(\overline{z} \mid z) > 0$, consider
  the updated policy transition $\kappa_1'$ that only differs from $\kappa_1$ in
  the following entries
  \begin{align*}
    \kappa_1'(z \mid \underline{z}) &= \kappa_1(z \mid
                                          \underline{z}) -
                                          (1-\alpha) t, \\
    \kappa_1'(\overline{z} \mid \underline{z}) &=
                                                     \kappa_1(\overline{z}
                                                     \mid
                                                     \underline{z})
                                                     + (1-\alpha) t, \\
    \kappa_1'(\overline{z} \mid z) &= \kappa_1(\overline{z} \mid
                                         z) - \alpha t, \\
    \kappa_1'(z \mid z) &= \kappa_1(z \mid z) + \alpha t,
  \end{align*}
  where
  \[
    \alpha = \frac{d(\underline{z})}{d(\underline{z}) + d(z)}
    \quad \text{and} \quad t =
    \begin{dcases}
      \kappa_1(z \mid \underline{z}), & \text{if}~\alpha = 0,\\
      \kappa_1(\overline{z} \mid z), & \text{if}~\alpha = 1,\\
      \min \left\{ \frac{\kappa_1(z \mid \underline{z})}{1-\alpha},
      \frac{\kappa_1(\overline{z} \mid z)}{\alpha} \right\}, & \text{otherwise.}
    \end{dcases}
  \]
  That is, we remove the $\underline{z} \to z$ transition if
  $d(\underline{z}) = 0$, remove the $z \to \overline{z}$
  transition if $d(z) = 0$, and make a compensated shift in the
  general case. In the first two cases, no individuals were affected
  by the policy transition. In the third case, $\alpha$ and $t$ are
  chosen so that $\kappa_1'$ is still a Markov kernel and $(1-\alpha)
  d(\underline{z}) = \alpha d(z)$.
  Thus,
  \begin{align*}
    & \E_{\kappa_1'}\{Y(Z')\} - \E_{\kappa_1}\{Y(Z')\}
    = \alpha d(z) t \cdot \left\{  - (\mu_{z} - \mu_{\underline{z}}) +
       (\mu_{\overline{z}} - \mu_{\underline{z}}) -
       (\mu_{\overline{z}} - \mu_{z})  \right\}
    = 0.
  \end{align*}
  So $\kappa_1'$ and $\kappa_1$ have the same average
  utility. Further, because $h$ is subadditive, we have
  \begin{align*}
    &\EH(\kappa_1', \gamma) - \EH(\kappa_1, \gamma) \\
    =& \alpha d(z) t \cdot \left[ - \E\{h(Y(z) - Y(\underline{z}))\} +
       \E\{h(Y(\overline{z}) - Y(\underline{z}))\} -
       \E\{h(Y(\overline{z}) - Y(z))\} \right] \\
    \leq& 0.
  \end{align*}
  Because the above is true for all $\gamma$, we have $\sup_{\gamma
    \in \Gamma} \EH(\kappa_1', \gamma) \leq \sup_{\gamma \in \Gamma}
  \EH(\kappa_1, \gamma)$, that is, $\kappa_1'$ can only have smaller
  worst-case harm.

  By making the above updates for finitely many times (use $\kappa_1'$
  as the new $\kappa_1$ and make the updates again), we arrive at a
  policy transition $\kappa_2$ that satisfies the first two properties of
  $\kappa^{*}$ in the Lemma statement and at least one of
  the following
  \begin{enumerate}
  \item[(i)] $\kappa_2(z \mid \underline{z}) = 0$ for all $\underline{z} < z$;
  \item[(ii)] $\kappa_2(\overline{z} \mid z) = 0$ for all
    $\overline{z} > z$.
  \end{enumerate}
  If (i) is true, we simply take $\kappa^{*} = \kappa_2$ so
  $\kappa^{*}$ satisfies all three desired properties (because all
  downward transitions have already been eliminated in Step
  1). Otherwise we move to the next step.

  \noindent \textbf{Step 3:} The final step is to remove all ``upward
  transitions'' to $z$. For each $\underline{z} <
  z$, define
  \[
    \beta_{\underline{z}} = (\mu_{z} - \mu_{\underline{z}}) /
    (\mu_{z^*} - \mu_{\underline{z}}).
  \]
  (If the denominator is zero, we simply take $\beta_{\underline{z}}
  = 0$.) Because $\underline{z} <
  z < z^*$, we have $0 \leq \beta_{\underline{z}} \leq 1$. Now consider the policy
  transition
  $\kappa_3$ that differs from $\kappa_2$ in the following entries:
  for all $\underline{z} < z$,
  \begin{align*}
    \kappa_3(z \mid \underline{z}) &= 0, \\
    \kappa_3(z^* \mid \underline{z}) &=
                                         \kappa_2(z^* \mid
                                         \underline{z}) +
                                         \beta_{\underline{z}}
                                         \kappa_2(z \mid
                                         \underline{z}), \\
    \kappa_3(\underline{z} \mid \underline{z}) &=
    \kappa_2(\underline{z} \mid \underline{z}) + (1 -
                                                     \beta_{\underline{z}})
                                                     \kappa_2(z \mid
                                                     \underline{z}).
  \end{align*}
  It is easy to see that $\kappa_3$ is still a Markov kernel and
  $\kappa_3(z \mid z) = 1$ (because Step 1 removed all downward
  transitions and $\kappa_2$ satisfies property
  (ii) above). The value of $\beta_{\underline{z}}$ is also chosen so
  that
  \[
     \E_{\kappa_3}\{Y(Z')\} - \E_{\kappa_2}\{Y(Z')\} = -
     d(\underline{z}) \kappa_2(z \mid \underline{z}) (\mu_{z}
     - \mu_{\underline{z}}) + d(\underline{z})
     \beta_{\underline{z}} \kappa_2(z \mid \underline{z}) (\mu_{z^*}
     - \mu_{\underline{z}}) = 0.
   \]
   So $\kappa_3$ has the same expected utility as $\kappa_2$.

   It remains to show that the worst-case harm of $\kappa_3$ is no
   greater than that of $\kappa_2$. Consider any $\gamma \in
   \Gamma$. Because $\Gamma$ is closed under monotone recoupling at
   $(z, z^{*})$,
   there exists $\gamma' \in \Gamma$ such
   that the marginal distribution of $(Y(0), \dots, Y(z - 1),
    Y(z+1),\dots,Y(M))$ is the same under
   $\gamma$ and $\gamma'$, and $Y(z) \leq Y(z^*)$ almost surely under
   $\gamma'$. % Consequently,
   % \[
   %   (Y(z) - Y(\underline{z}))_{-} \leq (Y(z^*) -
   %   Y(\underline{z}))_{-} \quad \text{almost surely for
   %     all}~\underline{z} < z.
   % \]
   Under this monotone recoupling, we have
   \begin{align*}
     &\EH(\kappa_3,\gamma') - \EH(\kappa_2,\gamma') \\
     =& \E_{\gamma'}
     \left\{\sum_{\underline{z} = 0}^{z - 1} - d(\underline{z})
       \kappa_2(z \mid \underline{z}) h(Y(z) -
       Y(\underline{z})) + d(\underline{z})
       \beta_{\underline{z}} \kappa_2(z \mid \underline{z})
        h(Y(z^*) - Y(\underline{z})) \right\} \\
     =& \E_{\gamma'}
     \left[ \sum_{\underline{z} = 0}^{z - 1} d(\underline{z})
       \kappa_2(z \mid \underline{z}) \cdot \left\{ - h(Y(z) -
       Y(\underline{z})) + \beta_{\underline{z}} h(Y(z^*) - Y(\underline{z}))
        \right\}  \right]\\
     \leq & 0.
   \end{align*}
   The last inequality is true because $\beta_{\underline{z}} \leq
   1$ and $h$ is nonincreasing.
   Moreover, the expected harm $\EH(\kappa_3,\gamma)$ only
   depends on the marginal distribution of $(Y(0), \dots, Y(z - 1),
    Y(z+1),\dots,Y(M))$ because
   $\kappa_3(z \mid z) = 1$. Thus,
   \[
     \EH(\kappa_3, \gamma) = \EH(\kappa_3, \gamma') \leq \EH(\kappa_2,
     \gamma').
   \]
   Because this is true for every $\gamma \in \Gamma$, we have
   \[
     \sup_{\gamma \in \Gamma} \EH(\kappa_3, \gamma) \leq \sup_{\gamma
       \in \Gamma} \EH(\kappa_2, \gamma).
   \]
   So by taking $\kappa^{*} = \kappa_3$, all three desirable
   properties of $\kappa^{*}$ are satisfied.
 \end{proof}

 \begin{proof}[Proof of \Cref{thm:benefit-equivalence-total-order}]
  Let $\mu_z(x)=\E\{Y(z)\mid X=x\}$ and write
  $\gamma_z^{-1}(\cdot\mid x)$ for the conditional quantile function
  of $Y(z)$ given $X = x$. The map
  $(y,y')\mapsto (y'-y)_+$ is submodular, so
  \citet[Cor.~2.3]{tchen_inequalities_1980} gives the pairwise
  comonotone minima. A common conditional $U\sim\operatorname{Unif}(0,1)$
  with $Y(z)=\gamma_z^{-1}(U\mid x)$ realizes all of them simultaneously.
  Under total stochastic ordering, the quantile difference has the sign
  of $\mu_{z'}(x)-\mu_z(x)$; equality of the means implies identical
  distributions. Hence, for every $\kappa$,
  \begin{equation}
    \inf_{\gamma\in\Gamma_{\mathrm F}}\EB_{\kappa,\gamma}
      =\E\left[\sum_{z,z'\in\mathcal Z}d(z\mid X)\kappa(z'\mid z,X)
        \{\mu_{z'}(X)-\mu_z(X)\}_+\right].
    \label{eq:robust-benefit-mean-positive}
  \end{equation}

  As in Step 1 of the proof of \Cref{lem:leapfrogging}, construct
  $\bar\kappa$ pointwise in $x$ by reallocating every off-diagonal
  transition $z\to z'$ with $\mu_{z'}(x)\leq\mu_z(x)$ to the source
  self-transition $z\to z$. In
  \eqref{eq:robust-benefit-mean-positive}, both the removed transition
  and its replacement contribute zero, so worst-case expected benefit
  is unchanged. For every coupling, the replacement also changes a
  nonnegative harm contribution into $h(0)=0$. Hence harm cannot
  increase and $\bar\kappa$ remains feasible.

  Every non-self transition retained by $\bar\kappa$ increases the
  conditional mean. We may therefore drop the positive part in
  \eqref{eq:robust-benefit-mean-positive}, obtaining
  \begin{equation}
    \inf_{\gamma\in\Gamma_{\mathrm F}}\EB_{\bar\kappa,\gamma}
      =\E_{\bar\kappa}\{Y(Z')\}-\E\{Y(Z)\}.
    \label{eq:oriented-benefit-gain}
  \end{equation}
  The same equality holds for $\kappa^*$ because it only stays at the
  current treatment or leapfrogs to a treatment with weakly larger
  conditional mean. By the mean-welfare optimality of $\kappa^*$ in
  \Cref{thm:lexical-leapfrogging},
  \begin{align*}
    \inf_{\gamma\in\Gamma_{\mathrm F}}\EB_{\kappa,\gamma}
    =\inf_{\gamma\in\Gamma_{\mathrm F}}\EB_{\bar\kappa,\gamma}
    =&\E_{\bar\kappa}\{Y(Z')\}-\E\{Y(Z)\} \\
    \leq&\E_{\kappa^*}\{Y(Z')\}-\E\{Y(Z)\}
    =\inf_{\gamma\in\Gamma_{\mathrm F}}\EB_{\kappa^*,\gamma}.
  \end{align*}
  Thus $\kappa^*$ solves \eqref{eq:opt-eb-eh}.
\end{proof}

\subsection{\texorpdfstring{Additional Theoretical Results and Proof of Results in \cref{sec:priority}}{Additional Theoretical Results and Proof of Results in ~\ref{sec:priority}}}\label{app:proofs-priority}

 \begin{prop}
\label{prop:quadratic}
We have
\begin{align*}
    \sup_{\gamma \in \Gamma(\theta_L)} \P\{Y(z^*) < Y(z)\}
    &= \tilde\gamma(z,\theta_L), \\
    \sup_{\gamma \in \Gamma(\theta_L)} \P\{Y(z^*) > Y(z)\}
    &= \tilde\gamma(z,\theta_L) - \E\{Y(z) - Y(z^*)\},
\end{align*}
where $\tilde\gamma(z,\theta_L) = \P(Y(z)=1, Y(z^*)=0)$ is the unique root of
\begin{equation}
    \label{eq:quadratic}
    (\theta_L - 1)\tilde\gamma^2 + \left[\theta_L(\gamma_{z^*} - \gamma_z) +
    \gamma_z + (1-\gamma_{z^*})\right]\tilde\gamma - \gamma_z(1-\gamma_{z^*}) = 0
\end{equation}
lying in the Fr\'{e}chet-feasible interval $[0,
\min(\gamma_z, 1-\gamma_{z^*})]$, given explicitly by
\begin{equation}
    \label{eq:root}
    \tilde\gamma(z,\theta_L) = \begin{cases}
        \gamma_z(1-\gamma_{z^*}) & \theta_L = 1, \\[6pt]
        \dfrac{-b(\theta_L) + \sqrt{b(\theta_L)^2 +
        4(\theta_L-1)\gamma_z(1-\gamma_{z^*})}}{2(\theta_L-1)}
        & \theta_L \in (0,1) \cup (1, \infty), \\[10pt]
    \end{cases}
\end{equation}
where $b(\theta_L) = \theta_L(\gamma_{z^*} - \gamma_z) + \gamma_z +
(1-\gamma_{z^*})$. In particular, $\lim_{\theta_L \to \infty} \tilde\gamma(z,\theta_L) = 0$ and $\lim_{\theta_L \to 0} \tilde\gamma(z,\theta_L) =  \min(\gamma_z, 1-\gamma_{z^*})$
\end{prop}

\begin{proof}
We will distinguish between the possible values of $\theta_L$ to determine the root of \cref{eq:quadratic}.

\textit{Case $\theta_L = 1$.} The leading coefficient vanishes and
\cref{eq:quadratic} reduces to
\begin{equation*}
    b(1)\cdot\tilde\gamma - \gamma_z(1-\gamma_{z^*}) = 0.
\end{equation*}
Since $b(1) = 1$, the unique solution is $\tilde\gamma(1) = \gamma_z(1-\gamma_{z^*})$.

\medskip
\textit{Case $\theta_L \in (0,1)\cup(1,\infty)$.} Let $Q(\tilde\gamma) =
(\theta_L-1)\tilde\gamma^2 + b(\theta_L)\tilde\gamma - \gamma_z(1-\gamma_{z^*})$,
and denote the lower Fr\'{e}chet bound of $\tilde{\gamma}$ by $L = \max(0, \gamma_z - \gamma_{z^*})$ and the upper by $U = \min(\gamma_z,
1-\gamma_{z^*})$. Since $z^*$ is the optimal treatment, $L=0$ and in this case $Q(0) = -\gamma_z(1-\gamma_{z^*}) \leq 0.$ For the upper bounds, we need to distinguish the two cases: when $U = \gamma_z$, we have
\begin{align*}
    Q(\gamma_z) &= \gamma_z\left[(\theta_L-1)\gamma_z +
    \theta_L(\gamma_{z^*}-\gamma_z) + \gamma_z + (1-\gamma_{z^*}) -
    (1-\gamma_{z^*})\right] \\
    &= \gamma_z \cdot \theta_L \gamma_{z^*} \geq 0;
\end{align*}
alternatively, $U = 1-\gamma_{z^*}$ and
\begin{align*}
    Q(1-\gamma_{z^*}) &= (1-\gamma_{z^*})\left[(\theta_L-1)(1-\gamma_{z^*}) +
    \theta_L(\gamma_{z^*}-\gamma_z) + \gamma_z + (1-\gamma_{z^*}) -
    \gamma_z\right] \\
    &= (1-\gamma_{z^*})\cdot\theta_L(1-\gamma_z) \geq 0.
\end{align*}

Since $Q(0) \leq 0 \leq Q(U)$, the intermediate value theorem guarantees exactly one root in $[0, U]$. To identify it, note that the two roots of
\cref{eq:quadratic} are
\begin{equation*}
    \hat\gamma^{\pm} = \frac{-b(\theta_L) \pm \sqrt{b(\theta_L)^2 +
    4(\theta_L-1)\gamma_z(1-\gamma_{z^*})}}{2(\theta_L-1)},
\end{equation*}
and their product is $\hat\gamma^+ \hat\gamma^- = -\gamma_z(1-\gamma_{z^*})/(\theta_L - 1)$.
When $\theta_L > 1$ this product is negative, so the roots have opposite signs
and the unique root in $[0, U]$ is $\hat\gamma^+$. When $\theta_L < 1$ the
product is positive and the sum $\hat\gamma^+ + \hat\gamma^- = -b(\theta_L)/(\theta_L-1) > 0$
shows both roots are positive; since $Q(0) \leq 0$ and the leading coefficient
is negative, the root in $[0,U]$ is the smaller one, which is again $\hat\gamma^+$.

\textit{Case $\theta_L \to \infty$.} Dividing \cref{eq:quadratic} by $\theta_L$ and taking
$\theta_L \to \infty$ we get:
\begin{equation*}
    \hat\gamma(\hat\gamma + \gamma_{z^*} - \gamma_z) = 0,
\end{equation*}
giving roots $\hat\gamma \in \{0, \gamma_z - \gamma_{z^*}\}$. Since
$\gamma_{z^*} > \gamma_z$, the only non-negative root is $\hat\gamma = 0 = L$,
corresponding to the comonotone boundary of the Fr\'{e}chet class.

\textit{Case $\theta_L \to 0$.} Setting $\theta_L = 0$ directly in
\cref{eq:quadratic} and dividing by $(-1)$:
\begin{equation*}
    \hat\gamma^2 - (\gamma_z + 1 - \gamma_{z^*})\hat\gamma +
    \gamma_z(1-\gamma_{z^*}) = (\hat\gamma - \gamma_z)
    (\hat\gamma - (1-\gamma_{z^*})) = 0,
\end{equation*}
giving roots $\hat\gamma \in \{\gamma_z,\, 1-\gamma_{z^*}\}$. Since
$U = \min(\gamma_z, 1-\gamma_{z^*})$, the unique root in $[0, U]$ is
$\hat\gamma = U$, corresponding to the countermonotone boundary of the Fr\'{e}chet class.
\end{proof}

\begin{lemma}\label{lem:neg-part}
    If $h\colon \R \to \R$ is subadditive, $h(y)=0$ for all $y \geq 0$ and $f\colon \R^2 \to \R$ defined by $f(x,y) := h(x-y)$ is sub-modular, then there exists a constant $c \geq 0$ such that
    \begin{equation*}
        h(y) = c\cdot y_-\qquad \text{for all }y \in \R.
    \end{equation*}
\end{lemma}

\begin{proof}
    To begin with, we show that $h$ is convex. To this end, let $x,y \in \R$ and assume $y \leq x$ without loss of generality. Since $f$ is sub-modular, it holds that
    \begin{equation*}
        h(a-b)+h(a'-b') \leq h(a-b')+h(a'-b)
    \end{equation*}
    for all real numbers $a \leq a'$ and $b\leq b'$. We now set
    \begin{equation*}
        a = \frac{x}{2},\qquad a' = \frac{x}{2} + \frac{x-y}{2},\qquad b = -\frac{y}{2}, \qquad b' = -\frac{y}{2} + \frac{x-y}{2}
    \end{equation*}
    and thus obtain
    \begin{equation*}
        h\left(\frac{x+y}{2}\right) \leq \frac{h(x)+h(y)}{2},
    \end{equation*}
    which proves that $h$ is mid-point convex. Moreover, $h$ is a non-increasing function because $h(b) = h(a+b-a) \leq h(a) + h(b-a) = h(a)$ for all $a \leq b$, which follows from the first and second assumption. Therefore, $h$ is also Lebesgue measurable and mid-point convexity is equivalent to convexity \citep{sierpinski_sur_1920}.

    Now, we can show that $h$ is a non-negative multiple of the negative part. To this end, let $y < x < 0$ and consider the ratios $\frac{h(y)}{y}$ and $\frac{h(x)}{x}$ in the following. First, we can use convexity of $h$ which yields
    \begin{equation}\label{eq:h-convex}
        h(x) = h\left(\frac{x}{y}\cdot y+(1-\frac{x}{y} )\cdot0\right) \leq \frac{x}{y}h(y) + (1-\frac{x}{y} )\cdot h(0) = \frac{x}{y}h(y),
    \end{equation}
    and thus proves that $\frac{h(x)}{x} \geq \frac{h(y)}{y}$.
    Second, we use that $h$ is subadditive. We can find $n \in\mathbb{N}$ and $r \in (x,0]$ so that $y = nx+r$ and iteratively apply subadditivity. Thus, we obtain
    \begin{equation*}
        h(y) \leq nh(x)+h(r) \leq n h(x) + \frac{r}{x} h(x) = \frac{y}{x} h(x),
    \end{equation*}
    where we used the same argument as~\eqref{eq:h-convex} for the second inequality. This shows that $\frac{h(x)}{x} \leq \frac{h(y)}{y}$ and combined with the first result that
    \begin{equation*}
        \frac{h(x)}{x} = \frac{h(y)}{y}
    \end{equation*}
    for all $y < x <0$. Since $h$ is non-increasing, we conclude that there exists a constant $c \geq 0$ such that $h(y) = c\cdot y_-$ for all $y \in \R$.
\end{proof}

\begin{proof}[Proof of Proposition~\ref{prop:copula}] First, we compute the expected harm under a fixed Gaussian copula. To this end, consider two random variables $Y(z^*)$ and $Y(z)$. By definition of $\Gamma(\rho_L)$, they have a Gaussian copula with parameter $\rho_z:= R_{z,z^*}$ and thus
    \begin{equation}\label{eq:copula}
        \begin{pmatrix}
        \yt_{z^*} \\ \yt_z
        \end{pmatrix} :=
        \begin{pmatrix}
        \Phi^{-1}\big(\gamma_{z^*}(Y(z^*))\big) \\ \Phi^{-1}\big(\gamma_z(Y(z))\big)
        \end{pmatrix}
        \sim \mathcal{N}\!\left(
        \begin{pmatrix}
        0 \\ 0
        \end{pmatrix},
        \begin{pmatrix}
        1 & \rho_z \\
        \rho_z & 1
        \end{pmatrix}
        \right).
    \end{equation}
    Since the conditional distribution of $\yt_z$ given $\yt_{z^*} = \ytl_{z^*}$ is known to be $
        \yt_z \mid \yt_{z^*}=\ytl_{z^*} \sim \mathcal{N}(\rho_z\ytl_{z^*}, 1-\rho_z^2)$, its cdf is given by
    \begin{equation*}
        \P(\yt_z \leq \ytl_z \mid \yt_{z^*} = \ytl_{z^*}) = \Phi\left(\frac{\ytl_z-\rho_z \ytl_{z^*}}{\sqrt{1-\rho_z^2}}\right),
    \end{equation*}
    and, plugging in $\ytl_z = \Phi^{-1}(\gamma_z(y_z))$ and $\ytl_{z^*} = \Phi^{-1}(\gamma_{z^*}(y_{z^*}))$, we obtain
    \begin{equation*}
        \P(Y(z) > y_z \mid Y(z^*) = y_{z^*}) = \Phi\left(\frac{\rho_z \Phi^{-1}(\gamma_{z^*}(y_{z^*})) - \Phi^{-1}(\gamma_z(y_z))}{\sqrt{1-\rho_z^2}}\right).
    \end{equation*}
    Using the tower property and the layer cake representation, we can now compute
    \begin{align*}
        \E[\{Y(z^*) - Y(z)\}_-]
        &= \E[\E[\{Y(z)-Y(z^*)\}_+\mid Y(z^*)]]\\
        &= \int_{-\infty}^\infty \int_{y}^\infty \P(Y(z) > s \mid Y(z^*) = y)\, \mathrm{d}s\, \mathrm{d}\gamma_{z^*}(y)\\
        &= \int_{-\infty}^\infty \int_{0}^\infty \Phi\left(\frac{\rho_z\, \Phi^{-1}(\gamma_{z^*}(y)) - \Phi^{-1}(\gamma_z(y+t))}{\sqrt{1-\rho_z^2}}\right)\, \mathrm{d}t\, \mathrm{d}\gamma_{z^*}(y).
    \end{align*}

    Next, we identify where the supremum is achieved. Consider two pairs of transformed potential outcomes $(\yt_{z^*}, \yt_z)$ and $(\yt_{z^*}',\yt_z')$ with correlation parameters $\rho_z \leq \rho_z'$. These two pairs of random variables can be stochastically ordered: by \citet[Ex.~9.A.8,~9.A.18]{shaked_stochastic_2007},  $(\yt_{z^*}, \yt_z) \leq_{\mathrm{PQD}} (\yt_{z^*}',\yt_z')$ which in turn implies $(\yt_{z^*}, \yt_z) \leq_\mathrm{sm} (\yt_{z^*}',\yt_z')$; here, $\leq_\mathrm{PQD}$ and $\leq_\mathrm{sm}$ denote the positive quadrant dependence and super-modular order, respectively. Since the $\yt$ are monotone transformations of the potential outcomes, we also have $(Y(z^*), Y(z)) \leq_\mathrm{sm} (Y'(z^*), Y'(z))$ by \citet[Thm.~9.A.9]{shaked_stochastic_2007}. Since the negative part is sub-modular, we have
    \begin{equation*}
        \E[\{Y(z^*) - Y(z)\}_-] \geq \E[\{Y'(z^*)-Y'(z))\}_-],
    \end{equation*}
    by definition of the super-modular order. Hence, the expected harm is non-increasing in $\rho_z$ and thus achieved at the smallest admissible value $\rho_L$.

    Lastly, we investigate the limit behavior. Since we have shown that the supremum is achieved at $\rho_L$, we consider a sequence of random vectors $(Y_{\rho_L}(z^*), Y_{\rho_L}(z))$ where $\rho_L \to -1$. By construction of the Gaussian copula, there exist two independent standard normal random variables $N^*$ and $N$ such that
    \begin{equation*}
        \begin{pmatrix}
            Y_{\rho_L}(z^*) \\ Y_{\rho_L}(z)
        \end{pmatrix} =
        \begin{pmatrix}
            \gamma^{-1}_{z^*}(\Phi(N^*))\\
            \gamma^{-1}_z(\Phi(\rho_L N^*+\sqrt{1-\rho_L^2}N))
        \end{pmatrix}.
    \end{equation*}
    By the continuous mapping theorem and because the inverse cdfs have only countably many discontinuity points,
    \begin{equation*}
        \begin{pmatrix}
            Y_{\rho_L}(z^*) \\ Y_{\rho_L}(z)
        \end{pmatrix} \to
        \begin{pmatrix}
            \gamma^{-1}_{z^*}(\Phi(N^*))\\
            \gamma^{-1}_z(\Phi(-N^*))
        \end{pmatrix} =
        \begin{pmatrix}
            \gamma^{-1}_{z^*}(U)\\
            \gamma^{-1}_z(1-U)
        \end{pmatrix}
    \end{equation*}
    almost surely, where $U$ is a uniform random variable. Applying the continuous mapping theorem again, we also obtain $\{Y_{\rho_L}(z^*) -Y_{\rho_L}(z)\}_- \to \{\gamma^{-1}_{z^*}(U) - \gamma^{-1}_z(1-U)\}_-$ almost surely. It remains to show that we can interchange the limit and the expectation. For this, we can use Vitali's convergence theorem which requires that the family of random variables $(\{Y_{\rho_L}(z^*) -Y_{\rho_L}(z)\}_-)_{\rho_L>-1}$ is uniformly integrable. To show this property, we use that $Y(z^*)$ and $Y(z)$ are integrable and that their marginal distributions are fixed and do not depend on $\rho_L$. For $M>0$, we perform the following estimations
    \begin{align*}
        \sup_{\rho_L \in (-1,1)}\E\Big[&\{Y_{\rho_L}(z^*) -Y_{\rho_L}(z)\}_- \cdot \,\mathds{1}\Big\{\{Y_{\rho_L}(z^*) -Y_{\rho_L}(z)\}_- > M\Big\}\Big]\\[1ex]
        &\leq \sup_{\rho_L \in (-1,1)} \E\Big[(\lvert Y_{\rho_L}(z^*) \rvert + \lvert Y_{\rho_L}(z)\rvert) \,\mathds{1}\Big\{\lvert Y_{\rho_L}(z^*) \rvert + \lvert Y_{\rho_L}(z)\rvert > M\Big\} \Big]\\[1ex]
        &\leq 2\E\Big[\lvert Y(z^*)\rvert \,\mathds{1}\Big\{\lvert Y(z^*) \rvert > \frac{M}{2}\Big\}\Big] + 2\E\Big[\lvert Y(z)\rvert \,\mathds{1}\Big\{\lvert Y(z) \rvert > \frac{M}{2}\Big\}\Big] \\[1ex]
        &\to 0\qquad \text{as } M\to \infty,
    \end{align*}
    which prove uniform integrability and conclude the proof.
\end{proof}

\begin{proof}[Proof of Corollary~\ref{prop:gaussian}]
Fix $x$ and suppress the argument $x$. For each $z$, stochastic
dominance implies
\[
  \Phi\!\left(\frac{t-\mu_{z^*}}{\sigma_{z^*}}\right)
  \leq
  \Phi\!\left(\frac{t-\mu_z}{\sigma_z}\right),
  \qquad t\in\R.
\]
Since $\Phi$ is strictly increasing, the coefficient of $t$ in the
resulting affine inequality must vanish. Hence
$\sigma_z=\sigma_{z^*}=\sigma$ and the remaining inequality gives
$\mu_z\leq\mu_{z^*}$.

Let $\tau=\mu_{z^*}-\mu_z$, $r=\tau/\sigma$, and
$a_L=\sqrt{2(1-\rho_L)}$.
For a coupling in $\Gamma(\rho_L)$, let $\rho\geq\rho_L$ denote the
correlation between $Y(z^*)$ and $Y(z)$. Then
\[
  W=Y(z^*)-Y(z)
  \sim \mathrm{N}\!\left(\tau,\,2\sigma^2(1-\rho)\right).
\]
By \Cref{prop:copula}, the supremum of $\E(W_-)$ is attained at
$\rho=\rho_L$. Define the standard normal loss function
$\ell(s)=\varphi(s)-s\Phi(-s)$, so for $W\sim\mathrm{N}(\mu,\sigma^2)$ we
have $\E(W_-) = \sigma\,\ell\!\left(\mu/\sigma\right)$.
Therefore, we have
\[
  \sup_{\gamma\in\Gamma(\rho_L)}
  \E_\gamma\{Y(z^*)-Y(z)\}_-
  =
  \sigma a_L\,\ell(r/a_L).
\]
From this we obtain the priority score in the Corollary.
Finally, using $\ell'(s)=-\Phi(-s)$, we obtain
\[
  g'(s)
  =
  \frac{\ell(s)-s\ell'(s)}{\ell(s)^2}
  =
  \frac{\varphi(s)}{\ell(s)^2}
  >0.
\]
Thus $g$ is strictly increasing. So when $\rho_L$ is common across
subgroups, the priority ranking is equivalent to ranking by $r(z,x)$.
\end{proof}
\section*{Acknowledgments}
  The authors thank Mats Stensrud and Sina Akbari for helpful comments on an earlier version of this work.
\section*{Funding}
  M.S. was supported by the Economic and Social Research Council through the Doctoral Training Partnership (grant ES/P000738/1). T.F. was supported by the Swiss National Science Foundation (grant 200021/207436).
\section*{Data and code availability}
Replication code and materials are available at
\url{https://www.statslab.cam.ac.uk/~qz280/publication/counterfactual-policy/}.

\bibliographystyle{plainnat}\bibliography{references}

\end{document}